\documentclass[times, twocolumn, twocolappendix]{aastex63}
\usepackage{graphicx, mathptmx}
\usepackage{advdate}

\usepackage[export]{adjustbox}
\newcommand{\msun}     {\ensuremath{{\mathit{M}}_{\scriptscriptstyle \odot}}}
\newcommand{\subsun}     {\ensuremath{_{\scriptscriptstyle \odot}}}
\newcommand{\kms}      {\ensuremath{\mathrm{km~s^{-1}}}}

\newcommand{\units}[1]  {\ensuremath{\mathrm{{#1}}}}

\def\gapprox{\ensuremath{\ga}}  \def\lapprox{\la}

\received{January 17, 2024}
\revised{July 10, 2024}
\accepted{July 22, 2024}
\submitjournal{AAS Journals}
\shorttitle{$M_{\mathrm{BH}}$ in M64}
\shortauthors{G{\"u}ltekin et al.}
\begin{document}

\title{The Black Hole Mass and Photometric Components of NGC 4826}

\correspondingauthor{Kayhan G{\"u}ltekin}
\email{kayhan@umich.edu}

\author[0000-0002-1146-0198]{Kayhan G{\"u}ltekin}
\affiliation{Department of Astronomy, University of Michigan, 1085 S University Ave, Ann Arbor, MI 48109, USA}

\author[0000-0002-8433-8185]{Karl Gebhardt}
\affiliation{Department of Astronomy, University of Texas, Austin, Texas 78712}

\author[0000-0001-9854-5217]{John Kormendy}
\affiliation{Department of Astronomy, University of Texas, Austin, Texas 78712}
\affiliation{Max Planck Institute for Extraterrestrial Physics, 
    Giessenbachstrasse, D-85748 Garching by Munich, Germany}

\author[0000-0002-1616-1701]{Adi Foord}
\affiliation{Department of Physics, University of Maryland Baltimore County, 1000 Hilltop Cir, Baltimore, MD 21250, USA}

\author[0000-0001-7179-0626]{Ralf Bender}
\affiliation{Max Planck Institute for Extraterrestrial Physics, 
    Giessenbachstrasse, D-85748 Garching by Munich, Germany}
\affiliation{University Observatory, Faculty of Physics, Ludwig-Maximilians-Universität München, Scheinerstr. 1, 81679 Munich, Germany; rzoeller@mpe.mpg.de}

\author[0000-0003-3234-7247]{Tod R.\ Lauer}
\affiliation{NSF National Optical Infrared Astronomy Research Laboratory, Tucson, AZ 85726, USA}

\author{Jason Pinkney}
\affiliation{Ohio Northern University, School of Science, Technology, and Mathematics, 525 S. Main Street, Ada, OH 45810}

\author{Douglas O.\ Richstone}
\affiliation{Department of Astronomy, University of Michigan, 1085 S University Ave, Ann Arbor, MI 48109, USA}

\author[0000-0002-0278-7180]{Scott Tremaine}
\affiliation{School of Natural Sciences, Institute for Advanced Study, 1 Einstein Drive, Princeton, NJ 08540}
\affiliation{Canadian Institute for Theoretical Astrophysics, University of Toronto,
60 St.\ George Street, Toronto, ON M5S 3H8, Canada}

\begin{abstract}

We present infrared photometry and \emph{Hubble Space Telescope} (\emph{HST}) imaging and spectroscopy of the Sab galaxy NGC 4826. Schwarzschild dynamical modeling is used to measure its central black hole mass $M$.  Photometric decomposition is used to enable a comparison of $M$
to published scaling relations between black hole masses and properties of host bulges.  
This decomposition implies that NGC 4826 contains classical and pseudo bulges of approximately equal mass.  The classical bulge has best-fit S{\'e}rsic index  $n = 3.27$.  The pseudobulge is made up of three parts, an inner lens ($n = 0.18$ at $r \lesssim 4\arcsec$), an outer lens ($n = 0.17$ at $r \lesssim 45\arcsec$), and a $n = 0.58$ S{\'e}rsic component required to match the surface brightness 
between the lens components.  The total $V$-band luminosity of the galaxy is $M_{VT} = -21.07$, the ratio of classical bulge to total light is $B/T \simeq 0.12$, and the ratio of pseudobulge to total light is $PB / T \simeq 0.13$.  The outer disk is exponential ($n = 1.07$) and makes up 
$D/T = 0.75$ of the light of the galaxy. 
Our best-fit Schwarzschild model has a black hole mass with $1\sigma$ uncertainties of $M = 8.4^{+1.7}_{-0.6} \times 10^{6}\ \msun$ and a stellar population with a $K$-band mass-to-light ratio of $\Upsilon_K = 0.46 \pm 0.03\ \units{\msun\ L_{\scriptscriptstyle\odot}^{-1}}$ at the assumed distance of 7.27 Mpc.  Our modeling is marginally consistent with $M = 0$ at the $3\sigma$ limit.  These best-fit parameters were calculated assuming the black hole is located where the velocity dispersion is largest; this is offset from the maximum surface brightness, probably because of dust absorption.  
The black hole mass---one of the smallest measured by modeling stellar dynamics---satisfies the
well known correlations of $M$ with the $K$-band luminosity, stellar mass, and velocity dispersion of the classical bulge only.
In contrast, the black hole is undermassive with respect to the correlation of $M$ with total (classical plus pseudo) bulge luminosity.  
Thus the composite (classical bulge plus pseudobulge) galaxy NGC 4826 is consistent with previous results on black hole scaling
relations and helps to strengthen these results at low black hole masses.
\end{abstract}

\keywords{Black holes (162), Galaxy bulges (578), Galaxy structure (622), Scaling relations (2031)}

\section{Introduction} 
\label{intro}
\object{NGC 4826} (M64) has been called the ``Evil Eye Galaxy'' \citep{1992Natur.360..442B}, the ``Black Eye Galaxy'' \citep{1993PASJ...45L..47V}, and the ``Sleeping Beauty Galaxy'' \citep{1994AJ....107..173R} on account of its prominent and beautiful dust lanes. 
Its morphological classification is Sab(s) \citep{1994cag..book.....S}, and it is an isolated galaxy with counter-rotating gas disks \citep{1992Natur.360..442B, 1993PASJ...45L..47V}.

Because of its proximity, NGC 4826 is an excellent galaxy to test ideas about low-mass black holes in galaxies.  
It is well established that black hole mass $M$ scales with host galaxy spheroid properties such as effective velocity dispersion \citep[$\sigma_e$;][]{2000ApJ...539L...9F, 2000ApJ...539L..13G} and luminosity \citep[$L_{\mathrm{bulge}}$;][]{1989IAUS..134..217D, 1993nag..conf..197K, 1995ARA&A..33..581K}.  In contrast, the scaling of $M$ with pseudobulge properties and total luminosity in disk galaxies is more uncertain \citep{2013ARA&A..51..511K, 2015ApJ...813...82R}.  Black holes in galaxies with pseudobulges tend to be undermassive compared to black holes in galaxies with similar-luminosity classical bulges \citep{2008MNRAS.386.2242H, 2010ApJ...721...26G, 2011Natur.469..374K,
2013ARA&A..51..511K, 2015ApJ...813...82R, 2016ApJ...817...21S}.
Also (see the above references), the distribution of their black hole masses is almost certainly the upper envelope of a distribution that extends to smaller black hole masses.
Understanding the evolution and demographics of low-mass black holes and of black holes in pseudobulges is important in part because
these black holes are closest in mass to the initial, ``seed'' black holes that grew via galaxy mergers and AGN-driving gas accretion
into the giant black holes that we see in giant classical bulges and elliptical galaxies. So, measurements of their masses put constraints 
on black hole formation models. 

The demographics of low-mass black holes are also important for space-based gravitational wave experiments such as the \emph{Laser Interferometer Space Antenna} (\emph{LISA}), which are most sensitive to gravitational wave frequencies produced by binary black holes with total mass $\sim10^{5}$--$10^{7}\ \msun$ \citep{2023LRR....26....2A}.  The number density of black holes in this mass range is poorly constrained because of the relative lack of direct dynamical mass measurements.  There are a number of  measurements of small black hole masses using maser observations \citep{2011ApJ...727...20K, 2016ApJ...826L..32G}.  The distribution of black hole masses from this sample pushes down to much smaller masses than is expected from their velocity dispersions and the $M$--$\sigma$ relation.  The undermassive black holes were shown by \citet{2013ARA&A..51..511K} to be either hosted by pseudobulges or in galaxies classified as ``mergers in progress''.  

The details of the evolution of black holes in pseudobulges are not well understood.  It has been thought that black holes can be fed through the same secular processes that make pseudobulges---the rearrangement of disk angular momentum by bars, globally oval disks, and
(in certain cases) global spiral structure.  These can form outer, inner, and nuclear rings; they can increase the central stellar mass
concentration in pseudobulges (thereby possibly destroying any bar), and they may feed AGNs (see \citealt{1993IAUS..153..209K, 2013seg..book....1K} and, e.g., \citealt{2004ARA&A..42..603K} for reviews). 
In galaxies that contain both a classical and a pseudo bulge (such as NGC 4826, section \ref{obs:imag:decomp}), the situation is likely more complicated.  In particular, we argue here that accretion of a gas-rich dwarf galaxy---not the more classic kind of secular evolution---may have built the galaxy's pseudobulge.  Cataloging the population of low-mass black holes and the population of black holes in pseudobulges 
is necessary to make progress.

In this paper we analyze NGC 4826 as a nearby galaxy with a small stellar bulge to probe the demographics of black holes in such hosts.  There is a wealth of space-based imaging and spectroscopic data to work with.
NGC 4826 was observed as part of \emph{HST} programs GO-8591 (PI: D.\ O.\ Richstone) with Wide Field Planetary Camera 2 (WFPC2) and Space Telescope Imaging Spectrograph (STIS).  It was also observed with the 
Hobby-Eberly Telescope Low Resolution Spectrograph.  We combined these data sets with public observatory databases for our analysis.

Section \ref{obs:imag} is a description of the imaging observations and photometric decomposition.  Section \ref{obs:spec} describes the spectroscopic observations and analysis required for modeling.  We describe our kinematic modeling and black hole mass estimation results of these data in section \ref{mod}.  We discuss our results, including the multiwavelength analysis of NGC 4826's nucleus and comparison of the black hole mass to scaling relations in section \ref{discussion}.  Finally, we summarize our results in section \ref{sum}.  

Throughout this paper we assume that the distance to NGC 4826 is $D = 7.27$ Mpc (as in \citealt{2013ARA&A..51..511K, 2013arXiv1308.6483K}) from surface brightness fluctuations 
(\citealt{2001ApJ...546..681T} corrected via \citealt{2009ApJ...694..556B}).  All distance-dependent quantities are scaled to this value.

{\subsection{Classical Bulges and Galaxy~Mergers;
Pseudobulges~and~Galaxy Secular Evolution}}
\label{obs:definitions}

{We emphasize that classical bulges and pseudobulges are defined
purely by observational properties and that our 
understanding of their evolution is separate from the
black hole detection and mass measurement presented in this paper.  However, it will turn out in Section
\ref{discuss:scaling} that our observational conclusions about correlations (or not) of
the NGC 4826 black hole with its host galaxy components will reinforce our
understanding of the relationships between morphological components
and galaxy evolution.}

{Classical bulges were originally defined \citep[e.g.,][]{1961hag..book.....S} as, in essence, elliptical galaxies that live in the middle of a galaxy disk.
Ellipticals (E) were characterized as having essentially elliptical isophotes, much higher central surface brightnesses than their associated disks plus steep brightness gradients, and old stellar populations.  At that time, it was thought that bulges and ellipticals
have surface brightness profiles $I(r)$ described by the \citet{1959HDP....53..275D} ``$r^{1/4}$ law'' that brightness depends on radius $r$ as 
$\log{I} \propto r^{1/4}$.  Modern work calls for refinement in this definition (see \citealt{2009ApJS..182..216K} for a review): both bulges and ellipticals have surface brightness profiles that are well described by \citet{1968adga.book.....S}  functions such that
$\log{I} \propto r^{1/n}$, where the S\'ersic index is in general different from the de Vaucouleurs value $n = 4$.  Experience (again see \citealt{2009ApJS..182..216K}) tells us the classical bulges and elliptical galaxies with absolute magnitudes $M_V$ \gapprox $-21.6$ mostly have S\'ersic index $1.8 \lapprox n \lapprox 4$, whereas ellipticals with $M_V \lapprox -21.6$  generally have $n \gapprox 4$.}  

{One further important addition is necessary for the definition of both ellipticals and classical bulges: in order to be so classified, they need to satisfy the E $+$ classical bulge ``fundamental plane'' structural parameter correlations between the effective radius $r_e$ that contains $1/2$ of the light of the component, effective surface brightness $\mu_e$ at $r_e$, velocity dispersion $\sigma$, and
absolute magnitude $M_V$
\citep[e.g.,][]{1987ApJ...313...59D, 1987nngp.proc..175F, 1987ApJ...313...42D, 1988ASPC....4..329D, 1992ASSL..178..337D, 1992ApJ...399..462B, 1993ApJ...411..153B}
or its projections
.  The reason is
that another kind of galaxy---usually called a ``spheroidal''
(Sph) or ``dwarf elliptical'' galaxy---also satisfies the
purely descriptive morphology of elliptical galaxies but proves to
have very different structural parameter correlations and, it turns out, also formation histories \citep[e.g.,][]{1984ApJ...282...85W, 1985ApJ...295...73K, 1987nngp.proc..163K, 2009ApJ...691L.142K, 2012ApJS..198....2K}.  This latter point is the reason why we
check our bulge--pseudobulge--disk decomposition parameters in the
Appendix.}

{We also need to emphasize that elliptical galaxies are
intrinsically much less flattened than disks; they have intrinsic
ellipticities of E0--E5 \citep{1970ApJ...160..831S, 1981MNRAS.194..679B}. Elliptical galaxies and 
classical bulges also show other features; e.g., the isophotes of elliptical galaxies can be distorted by a few percent in radius to be slightly ``more disky'' (equatorially flattened) or ``more boxy'' than exactly elliptical isophotes.  These details are not relevant here.  For the purposes of this paper, we can consider a classical bulge to be essentially indistinguishable from an elliptical galaxy of the same luminosity.}

{So far, we have said nothing about formation processes.  
However, galaxy evolution has been studied in detail, and those
studies add one more refinement to the above definitions.  A large body of work---observational, theoretical, and numerical simulation---has established with some confidence (although not without continuing debate) that elliptical galaxies and the classical bulges of disk galaxies formed via major mergers of 2 or more galaxies (or a series of such mergers) in which cold gas dissipation did not produce a disk component in the merger remnant.  By ``major'', we mean that the stellar mass ratio of the 
merger progenitors was $\gapprox 1/4$.  Either or both merger progenitors may have been an elliptical already or may have contained a classical bulge.  But the important refinement that helps us here is this: When the merger progenitor(s) contained a substantial outer disk component, the merger scrambles it into an elliptical galaxy with S\'ersic index $\sim 2$--$3$.
We know of no numerical simulation that produced an elliptical galaxy or classical bulge with $n \ll 2$.  We adopt this picture
of galaxy evolution and will use the result that classical bulges
do not have $n \ll 2$.  This paragraph summarizes a large volume of literature; the idea that mergers make ellipticals originates with \citet{1977egsp.conf..401T}, 
and a very partial list of reviews includes
\citet{1989ARA&A..27..235K, 1991ARA&A..29..239D, 1992ARA&A..30..705B}; and \citet{2016ARA&A..54..597C}.
A revealing numerical simulation in which the pre-existing disks in a major merger produce an elliptical galaxy with $n < 4$ is \citet{1994ApJ...437L..47M}.}

{The foregoing discussion of classical bulges is necessary in
order to understand the definition and importance of pseudobulges:}

{Whereas observations and galaxy-evolution theory progressed
together and influenced each other, it is important to realize
that the galaxy components now known as ``pseudobulges'' were
first recognized observationally
\citep{1979ApJ...227..714K, 1981seng.proc...85K, 1993IAUS..153..209K}.
Pseudobulges can be---and, for the purposes of this paper, are---defined purely observationally.  The classification criteria
that are most important here are that pseudobulges are more
disk-like than classical bulges.  In particular, they are flatter.  Also, rotation is more important in comparison to random motions than in classical bulges, and in many cases, their stellar velocity dispersions are smaller than the ridge line and scatter in the \citet{1976ApJ...204..668F} correlation between $\sigma$ and $M_V$.  Also, except in S0 galaxies, pseudobulges frequently contain cold gas and 
active star formation.  Criteria for recognizing
pseudobulges are listed in \citet{2004ARA&A..42..603K}; \citet{2011ApJ...733L..47F}; and \citet{2013seg..book....1K, 2016ASSL..418..431K}.  Pseudobulges that have been studied in
detail have brightness profiles that are well described by 
S\'ersic functions, with one classification criterion that we 
particularly use here: they can have S\'ersic indices $n \ll 2$.
Exponential ($n = 1$) brightness profiles or ones with even stronger outer cutoffs ($n < 1$) are common.}  

{Many authors have tried to 
classify pseudobulges using only one criterion; e.g., (1)
that $n < 2$ or (2) that they deviate from the \citet{1978bdcn.book.....K} correlation between 
$r_e$ and $\mu_e$, usually toward low surface brightness.  However, almost all of the half-dozen or so
classification criteria listed in the above references have
at least a few exceptions.  So it is safest to base classifications
on more than one criterion. Here, the first sign of a pseudobulge
contribution is apparent flattening; e.g., a local maximum in
ellipticity $\epsilon$ in Figure 6, accompanied by higher-surface-brightness ``ring'' in the brightness profile.  We also use the
constraint that classical bulges have $n$ \gapprox 2.}

{The studies that first recognized pseudobulges and that established the secular evolution paradigm concentrate on extreme examples---galaxies that contain essentially only a pseudobulge and a disk as
contrasted with galaxies that contain only a classical bulge and a
disk.  But these studies recognize that galaxies can
and often do contain both a classical bulge and a pseudobulge.
A few attempts to disentangle these components have been made \citep[e.g.,][and references therein]{2015MNRAS.446.4039E}.  NGC 4826
proves to be such a galaxy.  Moreover, the pseudobulge in NGC 4826
is complicated: it consists of two distinct disks
or rings with steep outer cutoffs ($n \ll 0.5$) plus a more diffuse part. The rings are easy to model definitively.  But the diffuse
part of the pseudobulge necessarily is difficult to distinguish from
the classical bulge.  Our photometric decomposition therefore
has coupled uncertainty between the classical bulge and part of the
pseudobulge.  For this reason, we show two ``error bar'' 
decompositions constrained by the known properties of pure classical bulges and pure pseudobulges.  
These uncertainties affect only Section 5.2, the comparison of
our derived black hole mass with bulge and pseudobulge total luminosity.
They do not affect our machinery to measure black hole mass.}

{We emphasize that our decomposition of the brightness distribution
of NGC 4826 into multiple classical bulge, pseudobulge, and
disk components is analogous to multi-Gaussian decompositions as 
used in black hole mass measurements
\citep[e.g.,][]{1992A&A...253..366M, 1994A&A...285..723E, 2002MNRAS.335..517V, 2002ApJ...578..787C}.
For purposes of mass measurement, it is not necessary to attach
physical interpretation to the components with which the stellar
density distribution is modeled.  It is necessary only that the
model reproduces the observed density distribution accurately.
In our case, bulge--pseudobulge--disk decomposition using S{\'e}rsic
functions allows us to do this accurately (Figure \ref{fig:decomp1r4}) with
fewer components than would be required with multi-Gaussian
expansion.  But interpretation of the components---which never
is relevant for multi-Gaussian expansion---is relevant here only
in Section \ref{discuss}.}

{Anticipating Section \ref{discuss}, we return to formation processes.  A large body of work has produced a new paradigm of the slow (``secular'')
evolution of disk galaxies that complements our picture of galaxy
evolution by hierarchical clustering and galaxy mergers.
Quoting \citet{2015HiA....16..316K}, ``Secular evolution happens because 
self-gravitating systems evolve toward the most tightly bound configuration that is reachable by the evolution processes that are available to them. They do this by spreading---the inner parts shrink while the outer parts expand. Significant changes happen only if some process efficiently transports energy or angular momentum outward.''
Common driving agents are galaxy bars and oval disk distortions.
Morphological features of galaxies such as nuclear, inner, and outer rings are now understood to be products of this secular evolution.  And pseudobulges appear to be a natural consequence of this evolution.
Observational and theoretical galaxy evolution work have proceeded in parallel and have illuminated each other.  However, the existence of
the above morphological features was revealed observationally
before they were understood to be part of the secular
evolution picture.  Secular evolution and its relation to galaxy
morphology is reviewed from an observational perspective by
\citet{1993IAUS..153..209K, 2013seg..book....1K, 2016ASSL..418..431K, 2004ARA&A..42..603K}; and from a theoretical perspective by \citet{1989dad..conf..231T} and \citet{2014RvMP...86....1S}.
Conferences that concentrate on secular evolution include
Special Session 3 on ``Galaxy evolution through secular processes''
at the 2012 IAU General Assembly \citep[see][]{2015HiA....16..316K}
and the 2011 Canary Islands Winter School on Secular Evolution \citep[see][]{2016ASSL..418..431K}.
Thus a substantial literature establishes the physical basis for
why and how pseudobulges form.  These results have proved to be useful in a variety of contexts.  For example, (1) every letter in a complicated galaxy classification such as (R)SB(r)b 
\citep{rc3}---where (R) means ``outer ring'' and (r) means ``inner
ring''---is now understood at least heuristically within the 
complementary paradigms of hierarchical clustering and secular galaxy evolution.  Also (2) correlations of bulge-like components with the masses of supermassive black holes are different for classical bulges and pseudobulges: Classical bulges correlate tightly with black hole mass, whereas pseudobulges show little or no correlation with black hole mass 
(see \citealt{2013ARA&A..51..511K} for a review and Section \ref{discuss:scaling} for NGC 4826 as an example).}

{We conclude this section with an important comment: Although we
identify and model the pseudobulge of NGC 4826 in the way that is done for galaxies that underwent secular evolution---that is, we define as a pseudobulge the inner parts of the galaxy that are diskier than the classical bulge---we will conclude that NGC 4826's ``pseudobulge'' was manufactured via a cold-gas accretion and not by evolution that was driven by a bar or oval.  To our knowledge, 
NGC 4826 is unique.  Its unusual evolution plausibly has been 
driven by gas that was accreted counter-rotating with respect to gas already in the galaxy.  The latter was evidently more important, 
because the gas in the dust disk co-rotates with the stars.  But interaction between co-rotating and counter-rotating gas will have
reduced the net angular momentum and dumped an unusual amount of gas---some of which formed the pseudobulge---near the center.  This is thought to be the reason for the unusually prominent dust disk.
} 

\section{Imaging and Photometry}
\label{obs:imag}

\begin{figure}[htb]
\includegraphics[width=\columnwidth]{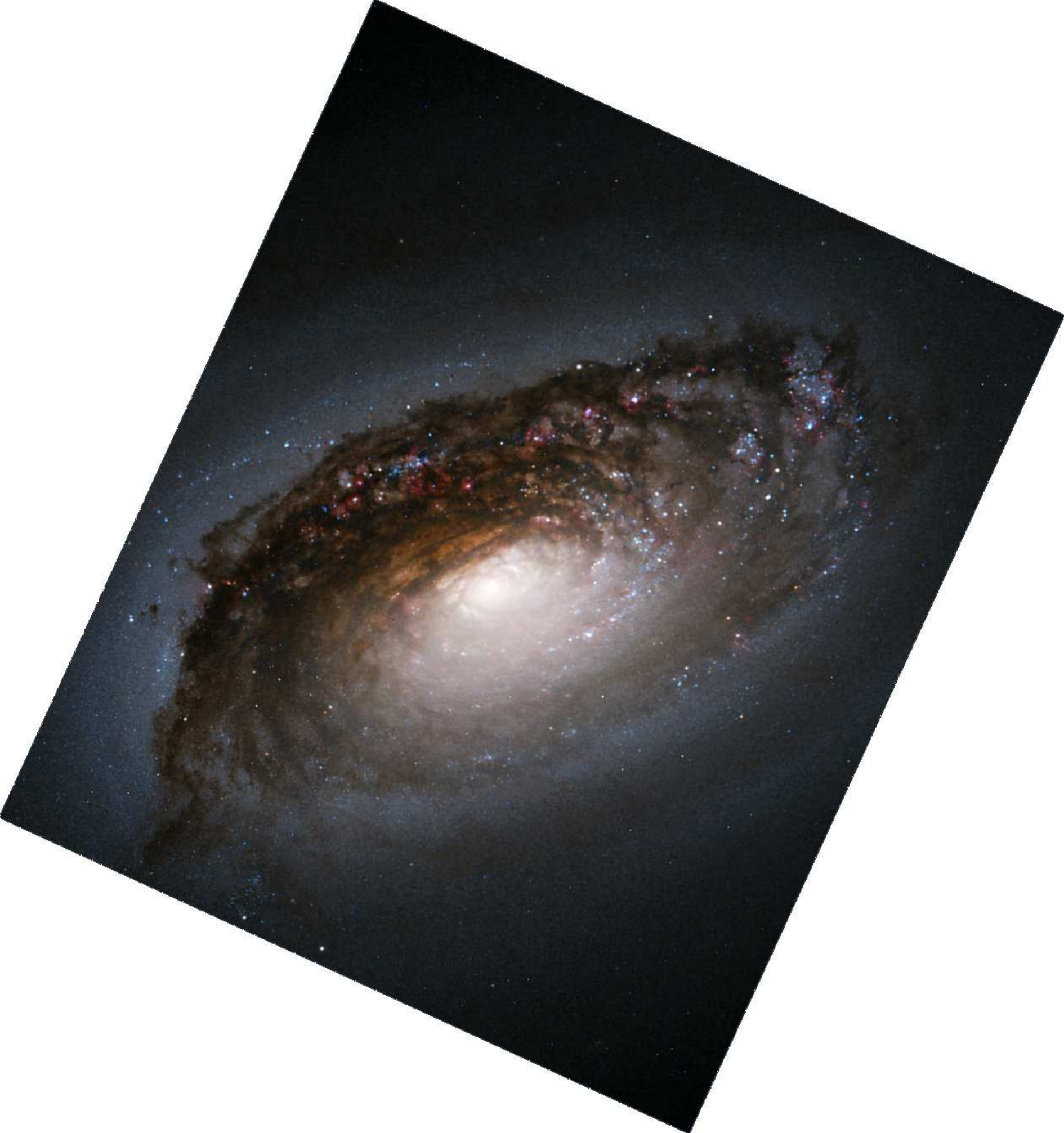}
\caption{Multiband image of NGC 4826 from STScI's Hubble Heritage project. \emph{HST} data to make the image come from two programs: GO-8591 and GO-9042. Colors are blue: F450W ($B$), cyan: F547M (Strömgren $y$), red: F656N (H$\alpha$), and pink: F814W ($I$).
{North is up; east is at left, and the NE side of the galaxy is the near side.  The long axis of the rectangular bounding box is 137\arcsec.}
The prominent dust lanes, a clear view of the small bulge, and inclination angle of the disk are evident from the image.  At small radii, there is a composite classical bulge plus pseudobulge \citep{2004ARA&A..42..603K,  2013seg..book....1K, 2013ARA&A..51..511K}.
Decoupling these components is necessary to interpret correlations between black hole mass and bulge properties.  Parameters of two overlapping components
are strongly coupled even in the absence of dust, and deriving them here required special care.  A key revelation from our analysis is that the internally absorbed and  very dark parts of the inner galaxy (in this figure) are 
actually higher in stellar density than the outer disk (Fig.\ \ref{fig:multicolorimage}).  To understand NGC 4826, we need 
photometry that is as absorption-corrected as possible.
\label{fig:image}}
\end{figure}

\begin{figure*}[htb]
\centering
\includegraphics[height=0.8\columnwidth]{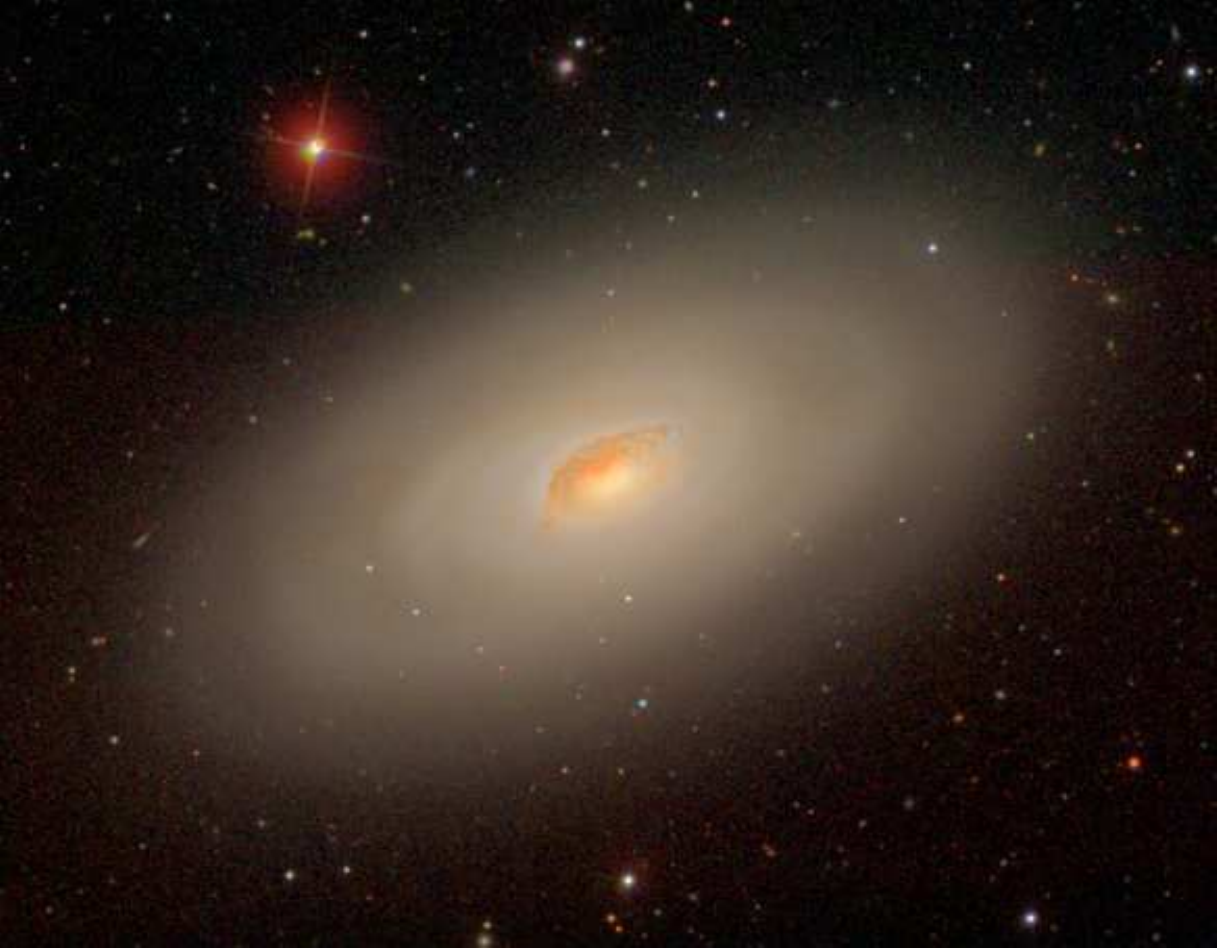}
\includegraphics[height=0.8\columnwidth]{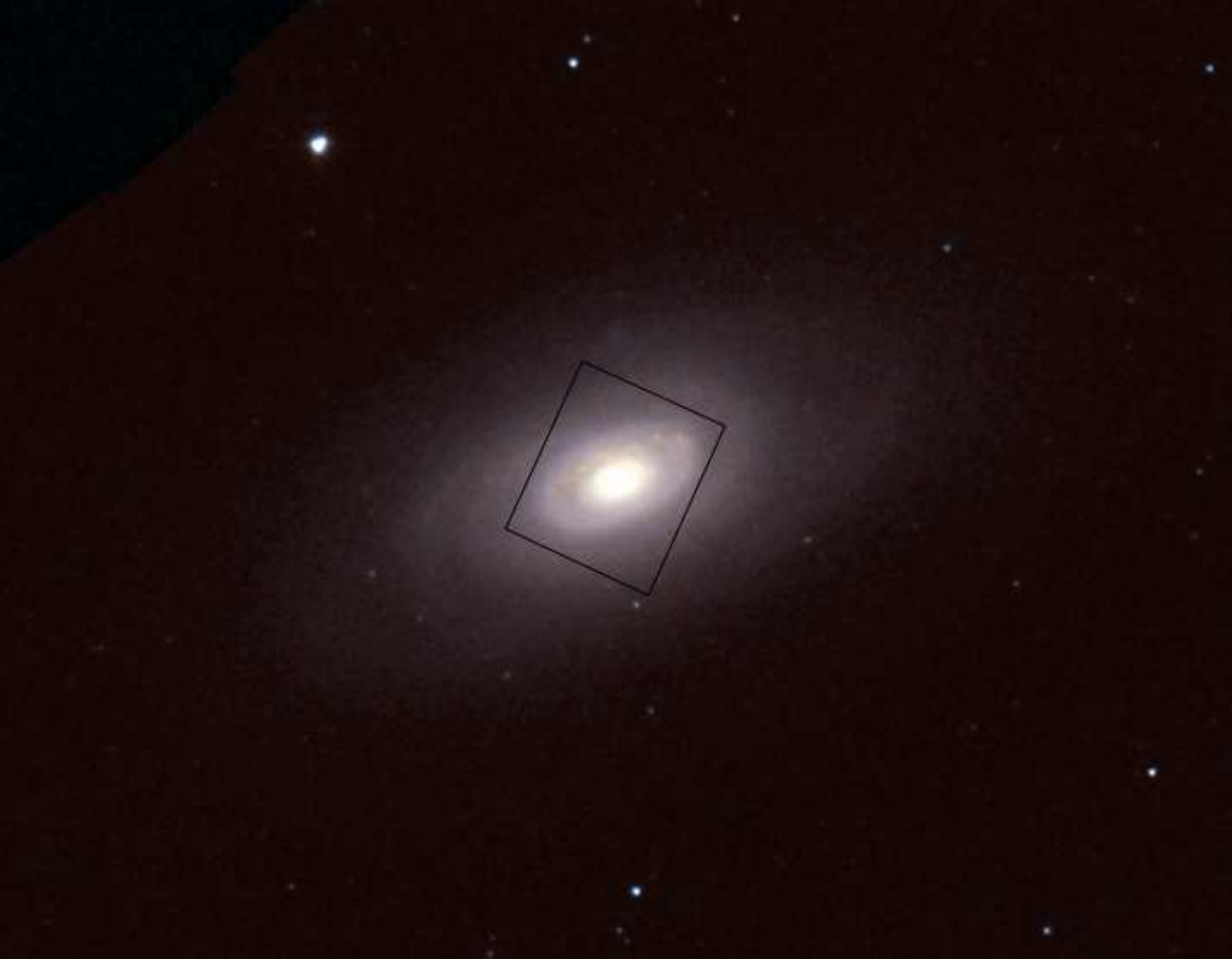}
\caption{Multicolor color composite images of NGC 4826.  The left image is a $gri$-band 
composite from the Sloan Digital Sky Survey via NED.  The right image is a color composite made
using the sum of 2MASS $J$, $H$, and $K$ images for the blue channel, the {\it Spitzer Space Telescope}
Channel 1 image at 3.55 $\mu$m for red and the average of the two for green.  Both panels show
the same field of view, the width of which is 11\farcm9; north is up and east is left.  The right panel shows the Figure \ref{fig:image} \emph{HST} Hubble Heritage field outlined
in black.  The dusty disk that dominates visible bandpasses is a slightly rim-brightened
bright disk (a ``lens''~component) when seen in the infrared, where dust extinction is less important.
\label{fig:multicolorimage}}
\end{figure*}

Figures \ref{fig:image} and \ref{fig:multicolorimage} show color composite images of the galaxy.  Figure \ref{fig:image} is the iconic \emph{HST} Hubble Heritage picture of the well known dust disk.  Figure \ref{fig:multicolorimage} shows, surprisingly, that the region that is strongly absorbed in Figure \ref{fig:image} is actually a bright ``shelf'' in the less-obscured infrared surface brightness and hence in stellar density.  This is classically called a ``lens'' component, although here it is more likely to be a result of star formation in accreted gas rather than a product of the different formation mechanism that makes lens components in barred galaxies \citep{2013seg..book....1K}. Figure \ref{fig:multicolorimage} emphasizes the lens component and the fact that the dusty lens region is only a small central part of a galaxy whose dominant disk extends to much larger radii.

\subsection{Near-Infrared \texorpdfstring{$K_s$}{Ks} Composite Profile Corrected for Internal Absorption}
\label{obs:imag:nir}
      We need to measure the near-infrared surface brightness profile corrected for internal absorption.  
Unlike the absorption in visible light (Figure 1), the absorption at $K_s$ is small enough to be
correctable, and the absorption at longer wavelengths $> 3\ \mu$m is assumed to be negligible.
This section provides non-parametric measures of surface brightness, ellipticity,
and position angle (PA) that are used for the stellar mass distribution in the dynamical modeling.
Separate from this non-parametric use, this profile is used in section \ref{obs:imag:decomp} to estimate the bulge, 
pseudobulge, and disk contributions of the galaxy.  We need to understand these components in order
to interpret a black hole mass ($M$) measurement or upper limit in the context of observed
correlations between $M$ and host galaxy component properties.

The $K_s$-band image that provides surface brightnesses and the zeropoint used in
this paper comes from the 2MASS Large Galaxy Atlas \citep{2003AJ....125..525J} via the NASA/IPAC Infrared Science
Archive\footnote{\tt https://irsa.ipac.caltech.edu/frontpage}.  The effective wavelength is 2.2 $\mu$m and the
scale is 0\farcs10 pixel$^{-1}$.  Foreground stars were cleaned by interpolation using a plane fitted to a
square annular region surrounding the star.  We were careful to exclude the PSF wings.
The image was unsmoothed at small radii and Gaussian smoothed with $\mathrm{FWHM} = 3$ pixels at
76$^{\prime\prime}$--165$^{\prime\prime}$ and
$\mathrm{FWHM} = 7$ pixels at larger radii.  Care was taken to ensure that smoothing does not degrade the profile.
Profiles were measured by JK using the \citet{1985ApJS...57..473L}  surface photometry package in the Lick Observatory image
processing system XVISTA \citep{1988igbo.conf..443S}.    All photometry discussed in
this section was carried out as above.

      In $K_s$ band, the dust disk is much less prominent than in the visible, but internal 
absorption remains significant.  We therefore corrected the region of the dust disk and smaller radii 
using the 2MASS $J$-band image and the procedure described in \citet{2008MNRAS.391.1629N} and \citet{2015ApJ...807...56B}.  Rephrasing the discussion from the latter paper for the present bandpasses, 
let $f_J$ and $f_{Ks}$ be the $J$- and $K_s$-band surface brightness fluxes per square arcsecond, respectively.  A subscript ``0'' refers to an extinction-corrected 
quantity, and the lack of it 
indicates magnitudes or fluxes as observed.  From the relation,
\begin{equation}
A_{Ks} \equiv K_{s}-K_{s,0} = \alpha E(J-K_{s}),
\label{eq:ak}
\end{equation}
where $A_{Ks}$ is the $K_{s}$-band absorption and $E(J - K_{s}) \equiv (J - K_{s}) - (J - K_{s})_0$ is
the reddening in the color $(J - K_{s})$, it follows that:
\begin{equation}
f_{Ks,{\kern 1pt}0} = \frac{f_{Ks}^{\alpha+1}}{f_J^{\alpha}}
\frac{f_{J,{\kern 1pt}0}^{\alpha}}{f_{Ks,{\kern 1pt}0}^{\alpha}}. 
\label{eq:fk}
\end{equation} 
We assume that the stellar population gradient in the inner part of NGC 4826 is negligible,
because~we see little color gradient between observed $H$ and $K$ profiles and between (see below)
dust-corrected $K_{s}$-band and observed 3.368 $\mu$m- and 3.55 $\mu$m-band profiles. Then
$f_{J,{\kern 1pt}0}/f_{Ks,{\kern 1pt}0}$ is approximately constant and so:
\begin{equation} 
f_{Ks,{\kern 1pt}0} \propto \frac{f_{Ks}^{\alpha+1}}{f_J^{\alpha}},  
\label{eq:fi}
\end{equation}
where 
\begin{equation}
\alpha = (A_J/A_{Ks} - 1)^{-1} \approx 0.62.
\label{eq:aj}
\end{equation}
Here we adopt the mean of the standard extinction curves in \citet{1998LNP...506..367F} and in \citet{2000asqu.book..523M}.
The correction is not perfect: it is based on the assumption that all of the dust is 
in a screen in front of the image, whereas here, most of the dust is expected to be in the
galaxy disk plane.  However, the correction works well, as evidenced by the
observation that the $\mu_{Ks,0}$ profile generated as above agrees with the profiles derived 
below in bandpasses with effective wavelengths $>3\ \mu$m.
In particular, the above extinction profiles give, in comparison to $V$ band,
$A_{Ks}/A_V \simeq 0.11$, $A_{\rm WISE}/A_V = 0.051$ ({\it WISE} observatory Band 1 at 3.368 $\mu$m),
and $A_{\rm Spitzer}/A_V = 0.050$ ({\it Spitzer Space Telescope} IRAC Channel 1 at 3.55 $\mu$m).
We neglect internal absorption in the latter two bandpasses.

      The central part of the adopted $K_s$-band composite profile was measured using two
images from \emph{HST} GO program 9360 (R.~Kennicutt, PI) and downloaded from the Hubble Legacy Archive\footnote{See \url{http://hla.stsci.edu/hlaview.html}.}.  Both were taken with the NICMOS NIC3 camera; the
scale is 0\farcs10 pixel$^{-1}$, and the filters used are F160W and F190N.  We tried to dust-correct
the F190N image using the F160W image, but the bandpasses are too  similar and the
procedure failed.  Noting that the maximum dust correction of the 2MASS $K_{s}$-band image was
only about 0.1 mag arcsec$^{-2}$, we assume that we get a good approximation to the absorption-free
near-infrared brightness profile of NGC 4826 by zeropointing the uncorrected F160W and F190N
profiles to the dust-corrected 2MASS profile where they overlap and where the resolution of 2MASS is 
still good enough.  The \emph{HST} profiles were zeropointed to 2MASS in this overlap region,
7$^{\prime\prime} \la r \la 28^{\prime\prime}$, where $r$ is major-axis radius.

      We get two important benefits from adding profile measurements from further into the infrared
at effective wavelengths 3.368 $\mu$m from the WISE observatory \citep{2010AJ....140.1868W} Band 1 and 
3.55 $\mu$m from the {\it Spitzer Space Telescope} \citep{2004ApJS..154....1W}  IRAC \citep{2004ApJS..154...10F} 
Channel 1, both downloaded from NASA/IPAC.  First, the {\it Spitzer} images 
have spatial resolution FWHM $\approx$ 1\farcs66 and (resampled)
scale 0\farcs75 pixel$^{-1}$ almost as good as that of 2MASS, but they are much deeper and they
have very constant sky brightnesses.  Thus we can measure the profile much farther out in the
galaxy with {\it Spitzer} than with 2MASS.  At wavelengths onger than $3\ \mu$m, the internal
absorption is also so small that it is effectively negligible.  Agreement between the {\it Spitzer}
$+$ WISE profiles and the dust-corrected (but not the dust-uncorrected) 2MASS profile gives us
confidence that we have removed the effects of the dust disk as well as possible.  The WISE
image has scale 1\farcs375 pixel$^{-1}$, but the resolution is $\sim6^{\prime\prime}$.
It is therefore useful only in the region of the dust disk, where it confirms the conclusions
from 2MASS and from {\it Spitzer} that what looks so faint in the optical is really a high-surface-brightness
disk component.  At $r < 16\farcs5$, the resolution of WISE is too poor to be useful, and at 
$r > 266^{\prime\prime}$, the galaxy surface brightness gets faint enough that confusion of
star images with each other becomes too serious to be corrected.  However, at intermediate radii, the 
agreement between the {\it Spitzer} and WISE profiles increases our confidence in the composite profile.
The {\it Spitzer} and WISE profiles are zeropointed to the dust-corrected 2MASS $K_s$ profile.

\begin{figure}[htb]
\includegraphics[width=\columnwidth,right]{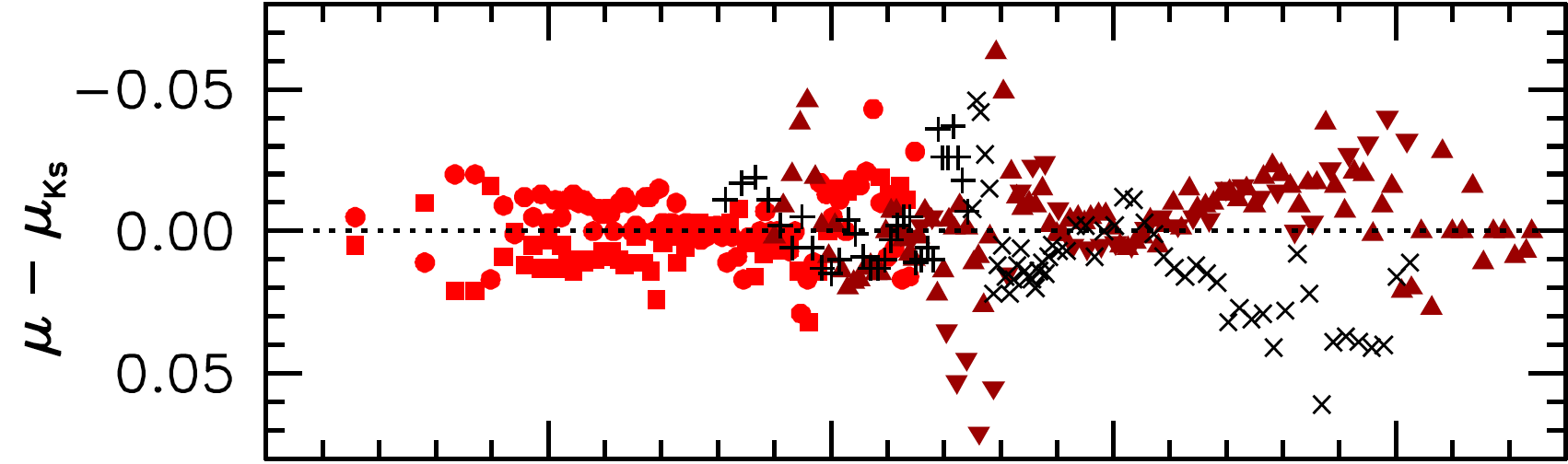}
\includegraphics[width=0.9825\columnwidth,right]{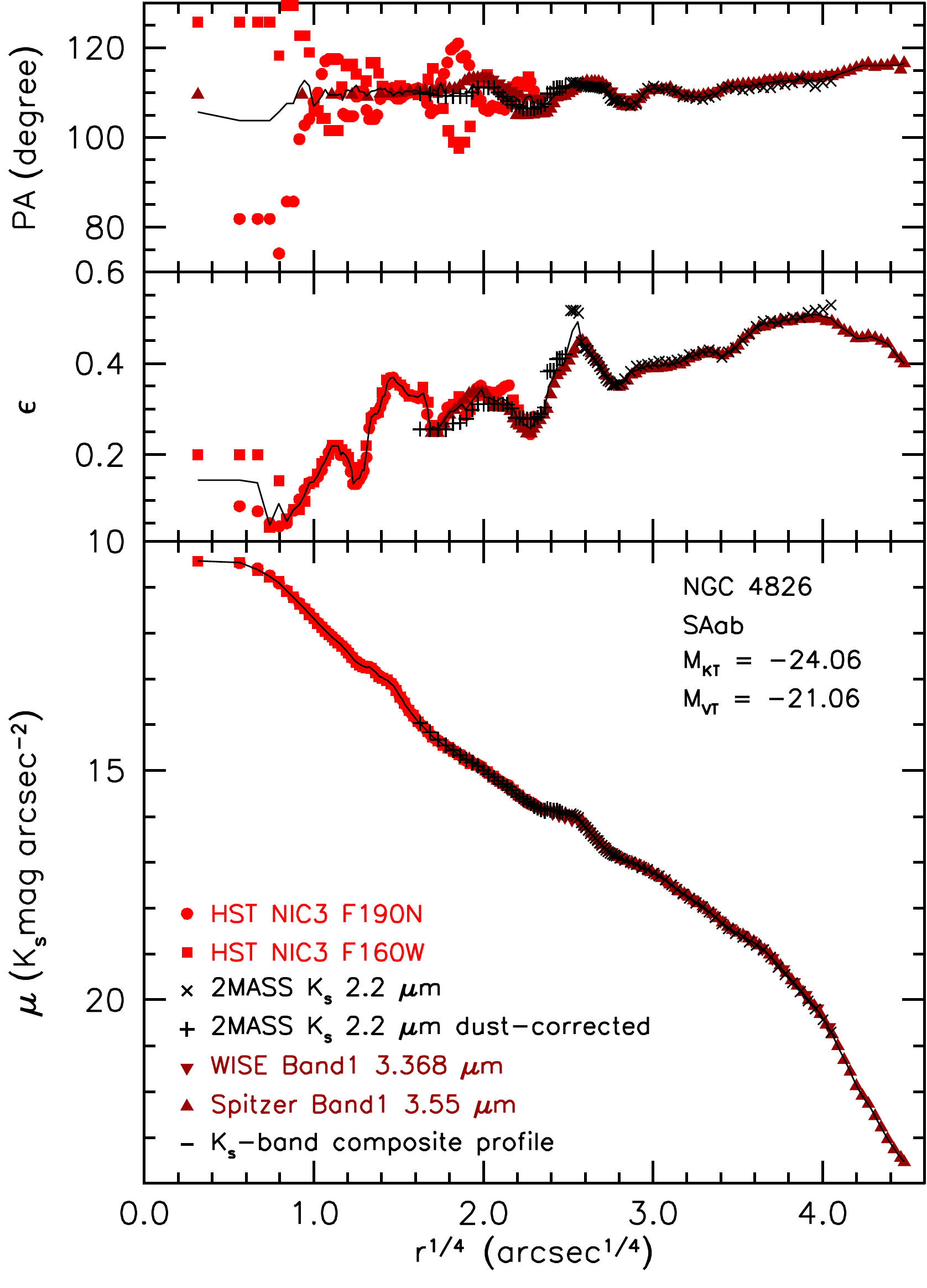}
\caption{
Composite surface photometry of NGC 4826 in $K_s$ and redder color bandpasses.  The $K_s$-band
zeropoint is on the 2MASS photometric system \citep{2massexplsup, 2000AJ....119.2498J}.
Outside the radii of the dust disk, the 2MASS $K_s$ image was measured without absorption corrections;
it provides the zeropoint.  Within the dust disk, the $K_s$ image was corrected for internal
absorption as described in the text.  The \emph{HST} NICMOS NIC3 profiles were then shifted to this 
absorption-corrected profile.  At large radii, 3.368 $\mu\mathrm{m}$ WISE and 3.55 $\mu\mathrm{m}$ {\it Spitzer Space
Telescope} profiles were shifted to the 2MASS $K_s$ profile.  The four panels show (bottom to top)
major-axis surface brightness, ellipticity, major-axis position angle PA east of north, and 
(given that the scatter is too small to see in the bottom panel) departures of the
individual profiles from the mean $K_s$ composite profile shown in the bottom panel by the solid
black curve.  It is important to note that the top panel shows that color gradients between $K_s$ and
$> 3\ \mu$m are small.  {Here and elsewhere, photometric profiles are plotted against $r^{1/4}$
so that S\'ersic $n < 4$ profiles (concave down) are easily distinguished from $n > 4$ profiles
(concave up).  We do not plot $\mu$ against $\log{r}$, because no components have $\mu \propto \log{r}$.}
\label{fig:kphot}}
\end{figure}

      All the above profiles are overplotted in Figure \ref{fig:kphot}.  Only the data that were averaged are
shown.  We pruned out points at small radii that are affected by spatial resolution that is poorer
than that of \emph{HST}.  We also pruned out points at large radii that are affected by non-uniform sky
brightness or by overlapping star PSFs.  The mean composite profile shown in Figure \ref{fig:kphot} by a black
solid curve is used in section \ref{obs:imag:decomp} to decompose the galaxy profile into classical bulge, disky
pseudobulge, and outer disk components.

      In the bottom panel of Figure \ref{fig:kphot}, deviations between the different telescope and bandpass
measurements are too small to be visible.  We therefore add a top panel which shows their 
deviations from the mean composite profile with an expanded magnitude scale.  The important
things to note here are (i) that the {\it Spitzer} and WISE profiles agree with the dust-corrected
2MASS $K_s$ profile---this provides reassurance about the profile measurements and dust correction---and (ii) that there is evidence for only a small color gradient, at least at these 
infrared wavelengths.  This justifies our assumption in Equation (\ref{eq:fi}) that the stellar population 
gradient is sufficiently small that we can assume that $f_{J,{\kern 1pt}0}/f_{K,{\kern 1pt}0}$ is approximately constant.

      Integrating the composite surface brightness and ellipticity profiles gives a total magnitude
of $K_{s,0} = 5.256$, where the subscript ``0'' indicates that the magnitude is corrected
for internal but not Galactic absorption.  The 2MASS Large Galaxy Atlas gives an integrated apparent 
magnitude of 5.330.  Given that 2MASS does not correct for internal absorption, this is excellent
agreement.

We adopt $(V - K_{s})_0 = 3.00$ from \citet{2013arXiv1308.6483K}.  Then the absolute magnitudes of the galaxy are $M_{K,{\kern 1pt}s} = -24.06$
and $M_V = -21.06$.  Component magnitudes are derived in Section \ref{obs:imag:decomp}.

\subsection{High-Resolution Central Imaging and Stellar Dynamical Model}
\label{obs:imag:hires}
To look for a supermassive black hole in NGC 4826, we need to construct dynamical models that have spatial
resolution much higher than those of either the photometry or the kinematic measurements.
Then models are constrained to fit the observations after PSF convolution.  This means that we need a
surface photometry model that is much higher in spatial resolution than the $K_s$-band photometry
achieved in Section \ref{obs:imag:nir} with the \emph{HST} NIC3 PSF and 0\farcs10 pixels.

\begin{figure}[htb]
\includegraphics[width=\columnwidth]{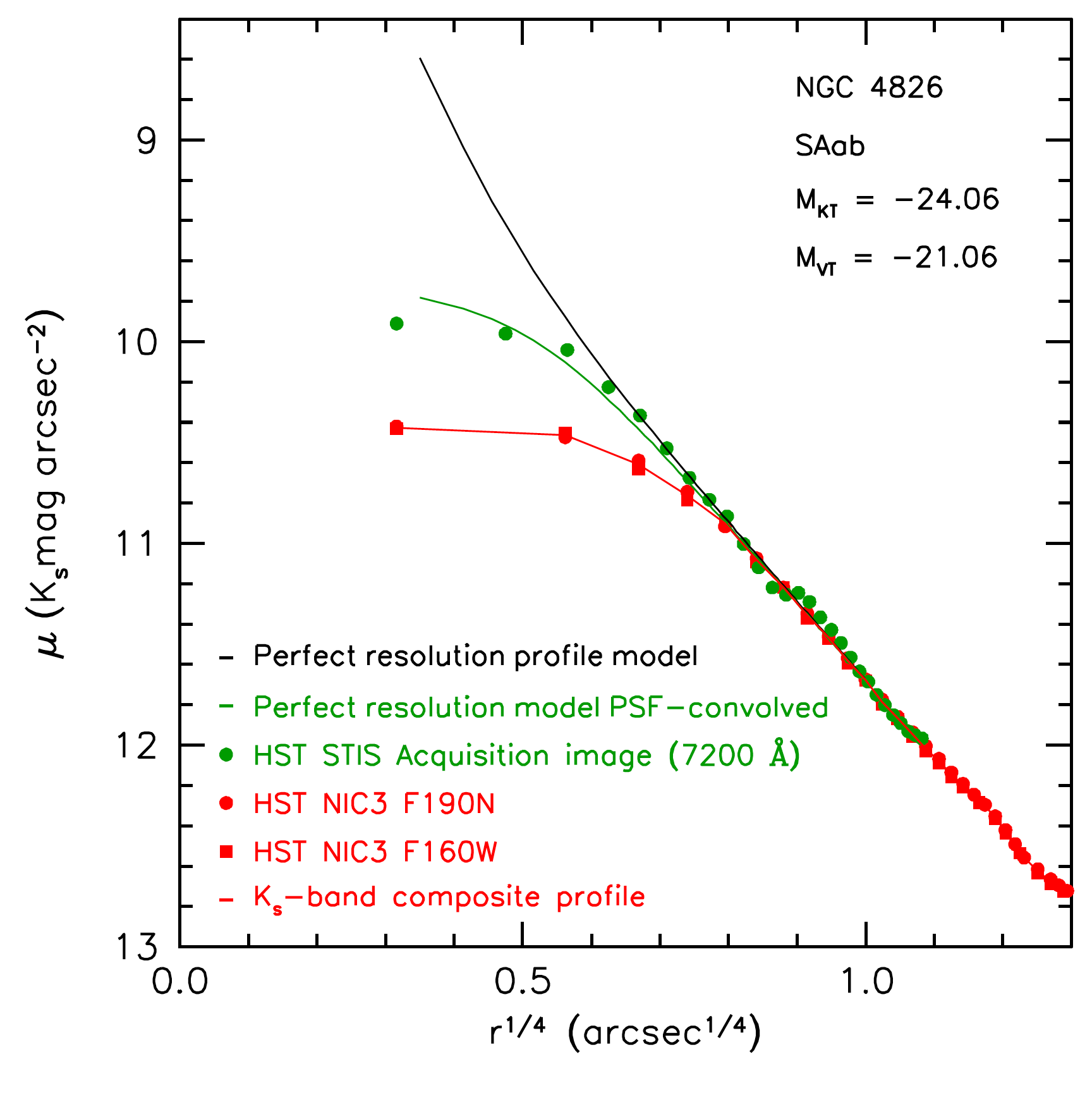}
\caption{
Photometric model of the stellar light distribution at small spatial scales used in dynamical modeling.  The NICMOS
NIC3 profiles and the inner part of the $K_s$-band composite profile from Figure 3 are shown in red.
A profile measurement of the 
$\sim7200\ $\AA\  
STIS acquisition image is shown with green points,
zeropointed to $K_s$.  The adopted perfect-resolution light profile used near the center of the galaxy 
is shown in black and (after PSF convolution) by the green curve.  At radii larger than those shown here, 
the $K_s$-band composite profile (including ellipticities) was used to model the stellar light distribution.}
\label{fig:photmodel}
\end{figure}

We used the STIS acquisition image to create a light model at the smallest scales.  STIS has a spatial
scale of 0\farcs0508 pixel$^{-1}$.  Figure \ref{fig:photmodel} shows the resulting profile zeropointed to $K_s$.  Also
shown is the adopted light profile at perfect spatial resolution (as used in the dynamical models)
and after convolution with a Gaussian PSF with $\mathrm{FWHM} = 0\farcs13$.  The unconvolved model surface brightness (black line) is able to match both the STIS and NIC3 data when convolved and sampled with the appropriate PSF and pixel scale.  Thus we are confident with the consistency of our multi-band photometry and ability to model the light distribution at small scales.

\subsection{Photometric Decomposition}
\label{obs:imag:decomp}
\begin{figure}[htb]
\centering
\includegraphics[width=\columnwidth]{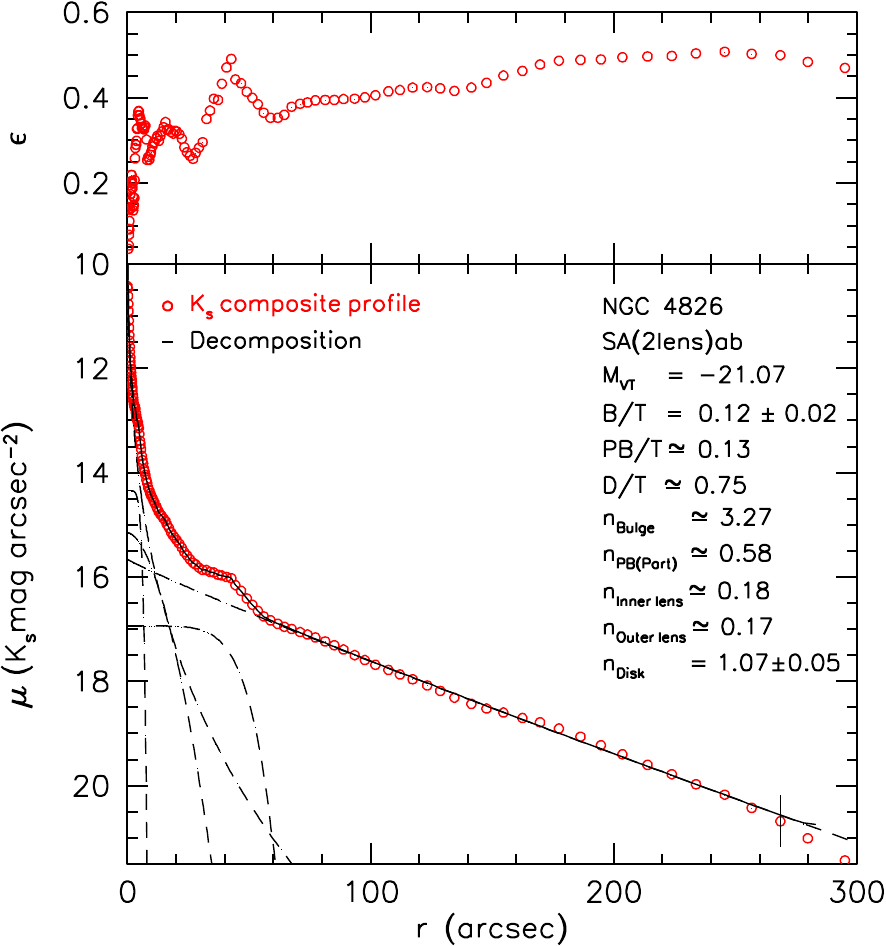}
\caption{
Adopted photometric decomposition of the composite $K_s$-band surface brightness profile of NGC 4826
here plotted against major-axis radius $r$ to emphasize that the galaxy is dominated in radius by its exponential 
disk.  The lens component revealed in Figure 2 is the prominent ``shelf'' in surface brightness with outer radius 
$\sim45^{\prime\prime}$.  A \citet{1968adga.book.....S} profile fit to this is shown by the outer dashed profile; 
it has index 
$n_{\rm Outer~lens} \simeq 0.17$ (with K-band surface brightness $\mu_{eK} = 16.75$ at effective radius $r_{eK} = 23\farcs0$). The inner components are, from the outside inward at 20 mag arcsec$^{-2}$, 
the outer lens, a 
classical bulge with $n_{\rm Bulge} = 3.27$ ($\mu_{eK} = 14.98$; $r_{eK} = 7\farcs4$), the
rest of the pseudobulge fitted with a S\'ersic function with $n_{\rm PB(Part)} = 0.58$ ($\mu_{eK} = 15.96$; $r_{eK} = 11\farcs3$), and an
inner lens with $n_{\rm Inner~lens} = 0.18$ ($\mu_{eK} = 14.38$; $r_{eK} = 3\farcs1$).  The disk is approximately exponential ($n = 1.07$ with K-band surface brightness $\mu_{eK} = 17.64$ at effective radius $r_{eK} = 101\farcs1$).
The inner components are better seen in Figure \ref{fig:decomp1r4}, which plots surface brightness against $r^{1/4}$.   The three parts of the pseudobulge (the outer lens, the rest of the pseudobulge [PB(Part)], and the inner lens) add up to have
$PB/T \simeq 0.13$, almost the same as the classical-bulge-to-total luminosity ratio $B/T \simeq 0.12 \pm 0.02$.
The sum of the components shown by dashed curves is the solid curve.  The decomposition fit is made between
the vertical dashes across the profile points (the inner one is at 0\farcs4).
\label{fig:decomp1r}}
\end{figure}

\begin{figure}[thb]
\centering
\includegraphics[width=\columnwidth]{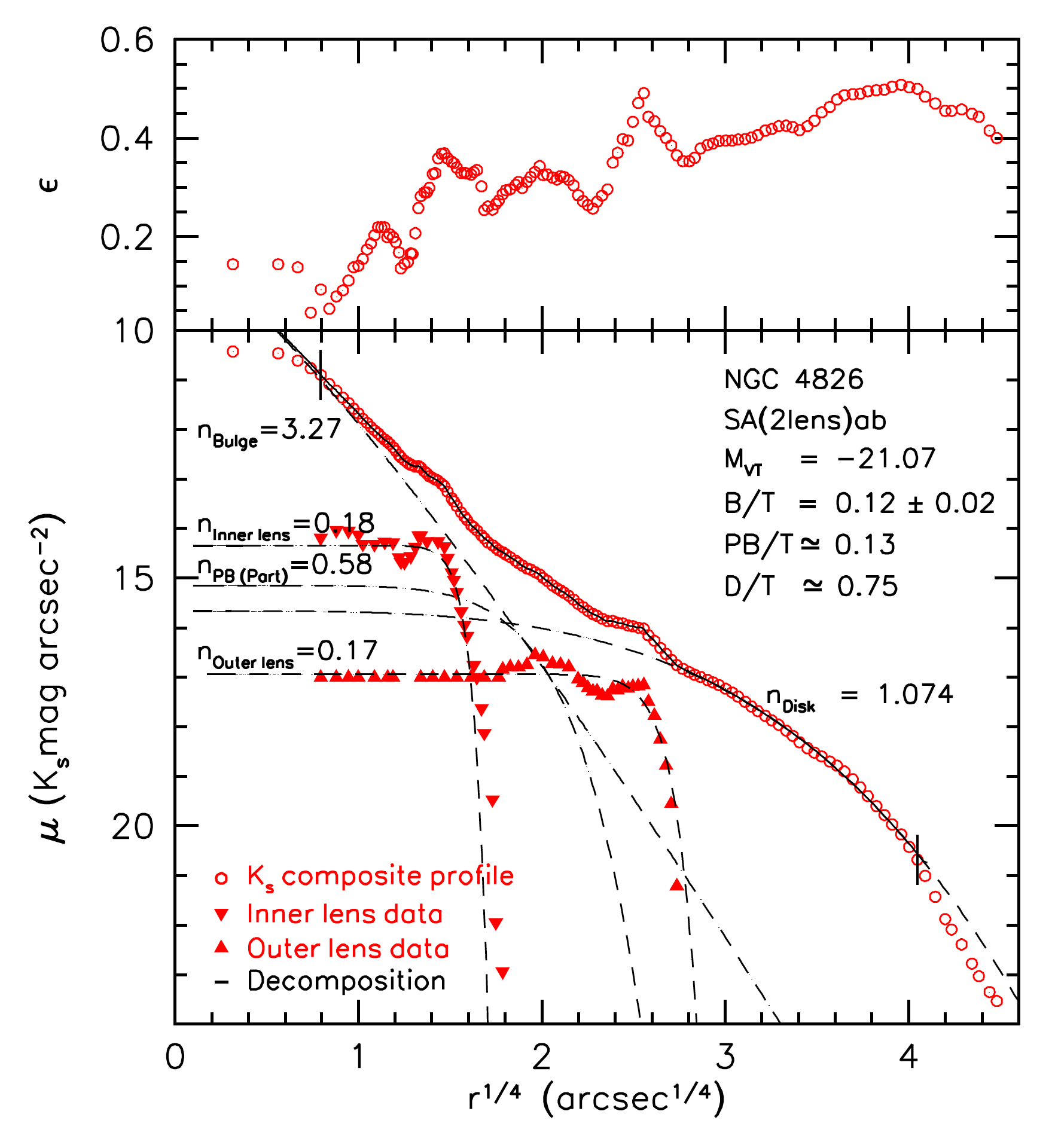}
\caption{
 Adopted photometric decomposition plotted against $(r/\mathrm{arcsec})^{1/4}$ to show components at both 
small and large radii.  The lens component shown in Fig.\ \ref{fig:multicolorimage} proves to be the outer of two ``shelves'' 
in surface brightness; these lens components have almost the same, very small S\'ersic index.  In practice,
we use their profiles as the data points that remain when the rest of the model is subtracted from the
composite profile.  Both sets of lens data points are chosen to have constant surface brightness inside
the radial range where they are well measured by the composite profile and decomposition model (that is:
we interpret these as lens components, not as rings that are dark inside).  When the
two lenses and the outer disk are subtracted from the composite profile, the rest of the galaxy is well
fitted by two S\'ersic functions, an inner one with $n_{\rm Bulge} \simeq 3.27$ that we identify as the
classical bulge and an outer one with $n_{\rm PB(Part)} = 0.58$ that we consider, together with the two
lens components, to be the pseudobulge.  The classical and pseudo parts of the bulge are approximately
equal in importance in this favored decomposition.  See Appendix and discussion in section \ref{obs:imag:decomp} for alternative decompositions.  Again, the sum of the components shown by dashed curves 
is the solid curve, and the fits are made between the vertical dashes across the profile points.
\label{fig:decomp1r4}}
\end{figure}

      Masses $M$ of supermassive black holes are observed to correlate with the luminosities,
stellar masses, and stellar velocity dispersions of elliptical galaxies and of classical bulges of disk galaxies
but not with galaxy disks, not with pseudobulges, and (apart from implications of the correlations with
ellipticals) not with dark matter halos (see \citealt{2013ARA&A..51..511K} for a review).  Therefore, to interpret 
the mass of any black hole that we detect in NGC 4826, we need to understand the structural components of the 
galaxy.  This requires photometric decomposition, 
and we use this decomposition as the basis of our surface-brightness deprojection to luminosity density, which is necessary for our kinematic modeling.
We adopt the photometric decomposition shown in Figs.\ \ref{fig:decomp1r} and \ref{fig:decomp1r4}.

Our understanding of the photometric decomposition possibilities in NGC 4826 is guided by previous studies of disk-dominated galaxies.
For reviews on classical bulges and ellipticals, see \citet{2009ApJS..182..216K} and \citet{2012ApJS..198....2K}; and for reviews of disk secular evolution and pseudobulges, see \citet{1993IAUS..153..209K, 2013seg..book....1K, 2016ASSL..418..431K} and \citet{2004ARA&A..42..603K}. Classical bulges and elliptical galaxies are essentially indistinguishable in their structure and structural parameter correlations.  At $M_V > -21.6$, they have S\'ersic indices $n < 4$; in fact, usually $n \simeq 1.8$ to 3 for small bulges like the one in NGC 4826.  Except  near the center (a few percent of the total light), classical bulges are  well described by single S\'ersic functions with slowly variable flattening, such that significant departures that are localized in radius or azimuth deserve interpretation.  The kind of pseudobulge that is relevant here\,---\,``disky pseudobulge''\,---\,is more disk-like than are classical bulges.  They are flatter (often as flat as the outer disk); and, in galaxies with overall star formation (not S0s), they generally contain disk gas and star formation. They often have smaller velocity dispersions than classical bulges.  A relevant observation for NGC 4826 is that the velocity dispersion decreases toward the center (cf.\ NGC 1553, whose pseudobulge has a smaller velocity dispersion than its surrounding disky lens component, \citealt{1984ApJ...286..116K}).  Disky pseudobulges are thought to form via star formation in gas that is driven toward galaxy centers by disk secular evolution --- by gradual rearrangement of the angular momentum distribution driven by galaxy bars, global oval distortions, and perhaps other global asymmetries (see the above reviews).  One reason for this idea is the  strong association of pseuduobulge classification features with the presence of such asymmetries.

In this context, it is important to emphasize that NGC 4826 shows no signs of any structural azimuthal asymmetries that are expected to drive secular evolution.  This does not  guarantee that such features did not exist in the past and have evolved away.  Bars destroy themselves if they increase the central mass concentration in a galaxy (see \citealt{2013seg..book....1K} for a review), e.g., via the above-mentioned secular evolution.  NGC 1553 is an example of a galaxy that now has a lens but no bar or discernible oval; the lens component may have been a bar in the past \citep{1979ApJ...227..714K, 1984ApJ...286..116K, 2013seg..book....1K}. We expect that lenses made via the azimuthal phase-mixing of elongated ``$x_1$'' bar orbits should have large radial velocity dispersions, and the lens of NGC 1553 has a large radial velocity dispersion \citep{1984ApJ...286..116K}.  The outer lens-like disk in NGC 4826, however, is cold.  There is strong evidence that NGC 4826 has accreted gas-rich small galaxies: the outer gas disk counter-rotates with respect to the inner galaxy. We see strong evidence for a pseudobulge contribution in NGC 4826.  This suggests that lens components specifically and disky pseudobulges more generally can be formed in another way that does not involve bar-driven secular evolution.  That is, pseudobulges can also form via gas accretion followed by star formation. Lenses formed this way are expected to be recognizable via their small velocity dispersions.  NGC 4826 proves to contain such a pseudobulge including two lens components.  The cause of the unusual structure of NGC 4826 plausibly is the recent accretion of gas that now counter-rotates with respect to the galaxy at large radii.  Collision of arriving, counter-rotating gas with corotating gas then still at large radii is expected to drive large amounts of gas toward the center \citep[e.g.,][]{1994AJ....107..173R, 1995ApJ...438..155R}.

Many galaxies are observed to contain only a classical 
bulge (e.g., M81) or only a pseudobulge (e.g., NGC 4736).  It is also true---and less 
well studied, because decomposition is tricky---that many galaxies contain both components \citep[e.g.,][]{2003ApJ...597..929E, 2013seg..book....1K, 2016ASSL..418..431K}.
NGC 4826 has long been considered to be in this category.  The weight of conflicting evidence led \citet{1993IAUS..153..209K}, \citet{2011Natur.469..374K}, and \citet{2013ARA&A..51..511K, 2013arXiv1308.6483K}
to classify the galaxy as having a pseudobulge.  Here, we recognize that both a classical and a pseudobulge are present, and we 
try to disentangle them to properly quantify their relative contributions.  Their parameters are
inevitably coupled.  As a result and based on experience summarized above, we derive three models,
a preferred one (Figs.\ \ref{fig:decomp1r} and \ref{fig:decomp1r4}) and two ``error bar'' models that have the smallest and largest
plausible classical bulges (Appendix Figs.\ \ref{fig:decomp2r4} and \ref{fig:decomp3r4}%
%See Appendix%
).  The error bar models are disfavored but not robustly disproved.  
Conclusions about black hole correlations are more robust if they are relatively independent of the decompositions on which they rely.

Note from the ellipticity profiles in Figs.\ \ref{fig:decomp1r} and \ref{fig:decomp1r4} that the outer lens component is, at its rim, 
as flat as the
flattest part of the outer disk.  Even the inner bump in the profile at $(r/\mathrm{arcsec})^{1/4} \simeq 1.5$ is flatter
than the surrounding (pseudo)bulge---it, too, is a disky ``inner lens'' component.  Also note that
the (pseudo)bulge is flatter between the two lens components than it is interior to the inner lens
component---it is not much thicker vertically than the thickest parts of the outer disk.  We conclude
that a disky pseudobulge contributes more importantly in this radial range $(r/\mathrm{arcsec})^{1/4} \simeq 2$ than it
does near the center.  Although the observation that $\sigma$ decreases toward the center in the STIS
spectroscopy leads us to suspect that some pseudobulge light contributes here, too.  It is worthwhile
to keep this in mind, but we do not include it in the decomposition.

      The preferred decomposition shown in Figs.\ \ref{fig:decomp1r} and \ref{fig:decomp1r4}  was derived iteratively, as follows:

      A first decomposition was made between 30$^{\prime\prime}$ and 260$^{\prime\prime}$, i.e.,
the radial ranges where the disk is exponential and, nearer the center, where the outer lens dominates
the light.  The disk was constrained to be nearly exponential; the S\'ersic index of the lens---which
included a contribution from the (pseudo)bulge---was a free parameter.  The decomposition program
converged to $n_{\rm Disk} = 1.07 \pm 0.05$ and $n_{\rm Outer~lens} = 0.17 \pm 0.03$.

      The above two fitted models were then subtracted from the composite profile and a decomposition
was run with a S\'ersic function fitted to the small bump in the profile at $(r/\mathrm{arcsec})^{1/4} \simeq 1.5$ 
(radius $\simeq 5^{\prime\prime}$) and another S\'ersic function fitted to the rest of the profile.
This gave a fit to what proves to be a tiny inner lens with $n_{\rm Inner~lens} = 0.18 \pm 0.03$.  This
fit is shown by a dashed line labeled with the S\'ersic index in Fig.\ \ref{fig:decomp1r4}, but at the end, it was replaced
with the data points shown by the upside-down filled triangles, i.e., the composite profile minus the
sum of the rest of the model fits (classical bulge plus pseudobulge plus outer lens plus disk).  Similarly,
the outer lens fit is shown by a dashed curve labeled with the S\'ersic index, but in the end, it was
replaced with the filled triangles, i.e., the composite profile minus the rest of the decomposition
model.  

      Finally, a profile was constructed from the $K_s$-band composite that should include only the 
classical bulge and the part of the pseudobulge that is not the inner or outer lens.  This profile was
decomposed into two S\'ersic components that are shown in Figs.\ \ref{fig:decomp1r} and \ref{fig:decomp1r4}.  The index of the classical
bulge, $n_{\rm Bulge} \simeq 3.27$ is a little larger than expected but is reasonable for a small bulge.
We have no expectations regarding the index of the rest of the pseudobulge; the decomposer chose 
$n_{\rm PB(Part)} \simeq 0.58$.  Then the classical bulge is 12\% $\pm$ 2\% of the light of the
galaxy, where the error bar reflects only the uncertainty in flattening, not any uncertainties about
whether this decomposition is preferred compared to the other two.  The sum of ``PB(Part)'' and the two lens
components is 13\% of the light of the galaxy; this agrees comfortably with the conclusion of earlier 
work (summarized in \citealt{2010ApJ...723...54K} and \citealt{2013ARA&A..51..511K}) that classical and pseudo components 
are roughly equally
important in this galaxy.  As noted above (e.g., in the context of the centrally decreasing $\sigma$),
if we have erred, it is most likely that we have underestimated the contribution of the
pseudobulge and overestimated the contribution of the classical bulge.

Note that, despite the Sab classification, the galaxy robustly consists mostly of a disk.  We find that 
the exponential outer disk has $D/T \simeq 0.75$ in this decomposition.  Moreover, the pseudobulge is also 
disky, even if its origin is different from that of the main disk.  We also do not, in the above inventory, include
any contribution from neutral or ionized gas; this is part of the disk, too.  NGC 4826 is a field galaxy; \citet{2010ApJ...723...54K}  emphasize that most giant galaxies in field environments are mostly or entirely
composed of disks.  They estimated $B/T \simeq 0.10$ and $PB/T \simeq 0.10$, consistent with the present, 
more detailed study, especially considering that, in 2010, we did not know about the outer lens
and other stellar light that is absorbed at visible wavelengths by dust.

Note again the diagnostic features in Figure \ref{fig:decomp1r4}: a disk with $D/T \simeq 0.75$ dominates the total
stellar light.  Its flattening varies with radius, perhaps reflecting a complicated formation history.
For example, the thickening at large radii may be associated with the accretion of the material that
now counter-rotates in the outer galaxy.  There are two well defined (albeit small: 1\% and 7\%
of the light of the galaxy) disky lens components; they, too, are highly flattened.  Classical and
pseudo bulges contribute roughly equally at radii between the two lenses; the galaxy is not much vertically
thicker there than it is in the inner exponential disk.  A classical bulge dominates only near the center, and
it contains only about 12\% of the stellar mass of the galaxy.

The formal uncertainties in Figs.\ \ref{fig:decomp1r} and \ref{fig:decomp1r4} and the subsequent interpretation underestimate true uncertainties in
the decomposition because the assumptions that we made are not encoded in uncertainties.  To get a better 
feel for the true uncertainties, we also constructed decompositions assuming the smallest and largest possible pseudobulges.  
Both were formally acceptable but at least some of their parameters would be outliers among all galaxies.  Figures of these 
alternative decompositions are in the Appendix.  Their construction is described here.

Minimal-pseudobulge model (see Appendix Fig.\ \ref{fig:decomp2r4}) was constructed by subtracting the two lens component data points from the total composite 
profile and making a two-S\'ersic decomposition of all the light that remains.  The resulting decomposition matches the observed profile in the fit range to within a few hundredths of a mag arcsec$^{-2}$.  The corresponding bulge-to-total ratio is $B/T = 0.30$, disk-to-total ratio is $D/T = 0.62$, and the lenses take up the remaining 8\% of the light.  This decomposition is disfavored for several reasons.  
The disk is not exponential: The best-fit $n_{\rm Disk} = 0.87 \pm 0.04$ is smaller than $n = 1$ for an exponential.  Also,
the S\'ersic \hbox{index} $n_{\rm Bulge} = 4.28 \pm 0.24$ of the classical bulge is bigger than experience leads us to expect for a 
small bulge.  Moreover, this decomposition ignores the implication of the ellipticity profile that a pseudobulge is 
important at more radii than just those of the two lenses. So this decomposition is inconsistent with experience with other galaxies.

The maximal-pseudobulge decomposition (See Fig.\ \ref{fig:decomp3r4} in the Appendix) was made by forcing the S\'ersic index of the bulge to be as small as possible.  
The best-fit decomposition
yielded parameters $n_{\rm Bulge} = 2.45 \pm 0.20$ to optimize the fit to the observed profile.  Then
$B/T = 0.09 \pm 0.01$.  An alternative solution is possible with $n_{\rm Bulge} = 1.93 \pm 0.19$, but it 
fits slightly less well, and it would imply the smallest S\'ersic index that we have seen for a classical
bulge.  Then we would have $B/T = 0.07$.  The maximal-pseudobulge fit, however, has smaller RMS($\mu$) and is adopted as our 
error bar decomposition with the smallest plausible classical bulge. It has $PB/T \simeq 0.16$ and $D/T \simeq 0.75$, 
similar to our adopted decompositions (Figs.\ \ref{fig:decomp1r} and \ref{fig:decomp1r4}).

All this may be complicated \hbox{enough~to~be~uncomfortable}, but many galaxies have composite classical plus pseudo bulge
central components plus, in the majority of cases, disks at large radii.  We need to be confident that our picture of the
correlations of black holes with host galaxies are not \hbox{biased} by any concentration on the simplest galaxies with the
fewest structural components and possibly the simplest formation histories.  So we should include galaxies
such as NGC 4826 in our samples. We need to be confident that our analyses---our photometric decomposition in particular here---leads to realistic results.  So we need to check that the parameters of the
classical bulge that we derive are realistic compared with the properties of simpler systems
with purely classical bulges and ellipticals with no disk component.  This is checked in Figures \ref{fig:FP} and \ref{fig:FJ} in
the Appendix. There, we see that the classical bulge of NGC 4826 is consistent with the projections of the 
fundamental plane structural parameter correlations.  The classical bulge of NGC 4826 is among the most compact
known, intermediate between M32 and more massive bulges but essentially normal.

Thus, we have a favored bulge-pseudobulge-disk decomposition and two extremal
decompositions to use in our comparison of NGC 4826 with published correlations between $M$
and host (pseudo)bulge properties.  
For the deprojection, we use the procedure due to {\citet{1996AJ....112..105G}}.  {The deprojection assumes axisymmetric spheroidal geometry of the luminosity density.  That is, the deprojection method assumes constant ellipticity, and the luminosity density is unique for a given inclination.  Assuming that the galaxy is axisymmetric but not spheroidal, i.e., with ellipticity that varies with radius, leads to non-unique deprojections and potentially large uncertainties for nearly edge-on systems \citep{1997MNRAS.287...35R}.  The ellipticities for each component were considered separately.  That is, the flattening of the spheroidal classical bulge was not assumed to be the same as the disky pseudobulge components.  As discussed in \citet{2000AJ....119.1157G}, the effect of incorrectly assuming spheroidal geometry will primarily lead to poorer constraints on $\Upsilon$ with little effect on the black hole mass estimate.}  

{For the case of NGC 4826, we assume that the bulge and disk are co-aligned so that measurement of the axis ratio of the disk (0.55) uniquely implies an inclination angle of $56^{\circ}$.}  To create our deprojected luminosity density, we use the five-component model that best describes the photometry.  For the lens components, we use the S\'{e}rsic fit, which neglects the small deviations seen in, e.g., Fig.\ \ref{fig:decomp1r4}, but these deviations are subdominant to the total surface brightness at their radii.
Each component is deprojected separately and then the luminosity densities are summed for the models. 
For all components (disk, lenses, partial pseudobulge, and classical bulge), we assume that the mass-to-light ratio and inclination are the same.  We note that our conclusions below about the mass of the central black hole are relatively insensitive to whether we include the lens and partial pseudobulges from our best-fit analysis versus our minimal- or maximal-pseudobulge model (see description above and figures in Appendix).  This is unsurprising as the lens and partial pseudobulge components are everywhere subdominant to the sum of the classical bulge and disk.  Even in the maximal-pseudobulge decomposition, the sum of the classical bulge and disk components is dominant everywhere but a small range of radii ($1.8 < (r/\mathrm{arcsec})^{1/4} < 2.0$), where the partial pseudobulge is comparable.  Nevertheless, it makes sense to use the best photometric decomposition as the source of our luminosity density.

\section{Spectroscopy}
\label{obs:spec}
Our \emph{HST} STIS spectra allowed us to extract kinematic information from both the gas emission lines (section \ref{obs:spec:halphaspec}) and stellar absorption lines (section \ref{obs:spec:cat}).  {We only use the kinematics from the stellar absorption lines for our modeling, but the gas velocity information is interesting in its own right as well as extremely helpful for determining the kinematic center of the galaxy (section \ref{obs:spec:kincen}.)}

\subsection{\texorpdfstring{H$\alpha$}{H-alpha} STIS Spectra}
\label{obs:spec:halphaspec}
The STIS spectrum centered on $\lambda = 6581\ $\AA\ (H$\alpha$ spectra) did not provide an independent black hole mass estimate, but it did provide useful information about the location of the black hole in the Ca II data.  This is because the position of the 0\farcs1 slit is the same for the two datasets; the centroid of the galaxy continuum falls on the same row in both.  

Our reduction for the STIS H$\alpha$ data began with a
pipeline-processed {\tt .flt} file. The processing includes bias subtraction, flat-fielding, and dark subtraction by the best reference dark.  The two CR-split exposures were summed while rejecting cosmic rays. Remaining cosmic rays were removed without clipping emission lines. The 2D spectrum was rectified using a rotation of $-0\fdg399$.  We fit our own wavelength solution from the associated {\tt .wav} file using air wavelengths.       

\begin{figure}[hbt]
\includegraphics[width=\columnwidth]{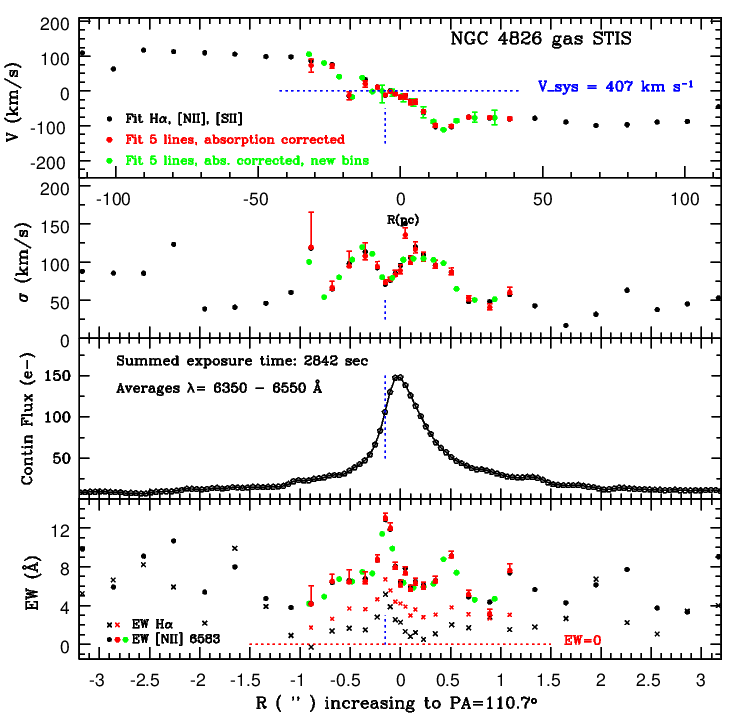}
\caption{
Kinematic parameters measured from H$\alpha$, [\ion{N}{2}], and [\ion{S}{2}] emission lines in STIS longslit spectroscopy.  From top to bottom, we show velocity, velocity dispersion, continuum flux, and equivalent width of H$\alpha$ and [\ion{N}{2}].  A vertical dotted line marks the likely dynamical center, which is notably offset from the location of the continuum maximum.
\label{jpgas}
}
\end{figure}

The kinematics of the ionized gas are shown in Figure \ref{jpgas}.
The five prominent emission lines were fit simultaneously for a common velocity and velocity dispersion.  Independent line strengths were solved for, except that [\ion{N}{2}] 6584 \AA\ was constrained to be 2.95 times stronger than 6548 \AA.  Emission lines were first fit in 48 bins extending symmetrically to $\sim$5.4\arcsec\ from the continuum peak.
The mean of these velocities provides a systemic velocity of
407.3 km s$^{-1}$.  This agrees to 1 \kms with published radio values \citep{deblok08}.
Next, emission lines were fit in only the inner 1\arcsec
after an absorption template was redshifted, broadened and
subtracted using the \ion{Ca}{2} kinematics as reference.
This has the main effect of boosting the H$\alpha$ line strength.  The $V$ and $\sigma$ measurements were not significantly changed due to absorption correction.  Finally, the green points in Fig.\ \ref{jpgas} are fits using a different binning scheme where the minimum bin size is 2 pixels.  We find that the kinematic trends discussed  below, and asymmetric $1\sigma$ error bars, are robust.

\subsection{Ca II STIS Spectra}
\label{obs:spec:cat}

The STIS observations of NGC 4826 were used for \ion{Ca}{2} triplet absorption measurement.  The spectrograph was operated with the G750M grating.  The $52\arcsec \times 0\farcs1$ slit was positioned at the center of the galaxy, aligned with the galaxy major axis as determined from ground-based images ($\mathrm{PA} = 110.^{\circ}3$).  Two exposures at central wavelength 8561\ \AA\ were taken at two dither positions for a total exposure time of 4190\ s, and a third exposure was taken at central wavelength 6581\ \AA\ for 2842\ s.  The wavelength range of the spectra targeting the \ion{Ca}{2} absorption was 8275--8842\ \AA. 

We followed standard pipeline procedures for our STIS data reduction.  First, we extracted the raw spectra from the dataset and then we subtracted a value coming from a constant fit to the overscan region for bias determination.  We used the \citet{2003ApJ...596..903P} iterative self-dark technique to take into account dark current as well as the warm and hot STIS CCD pixels.  Next we flat-fielded, dark-subtracted, and shifted the spectra to a common dither position to combine into a final two-dimensional spectrum.  One-dimensional spectra were then extracted using a bi-weight combination
of rows.

\subsection{Location of the Kinematic Center}
\label{obs:spec:kincen}
The photometric center of a galaxy normally has a velocity which matches the systemic velocity{, but this is not the case with NGC 4286.  As we outline below, examination of both the gas emission line profiles and the stellar absorption line dispersion profiles lead us to conclude that the kinematic center of the galaxy is offset from the photometric center, which we assume to be coincident with the peak continuum flux of the spectra.  The fact that both the gas profiles and the stellar profiles agree on this point gives us greater confidence in our kinematic center determination.}

{The gas velocities at the continuum peak are about 19$\pm$5 \kms\ below the systemic velocity (marked with a blue dotted line in the top panel of Fig.\ \ref{jpgas}).  The significance of this velocity offset is bolstered by the neighboring bins which are also below systemic and smoothly rise to 407 km s$^{-1}$ at $R\sim -0\farcs15$, which we mark with a vertical dotted line in all panels of Fig.\ \ref{jpgas}.  The offset of the continuum peak is likely due to a gradient in dust absorption, and the third panel of Fig.\ \ref{jpgas}
confirms that the continuum intensity drops off more steeply
at $R < 0$ than $R > 0$.  The second panel shows a dip (rather than a peak) in the gas velocity dispersion at $R \sim -0\farcs15$, but the overall gull-wing profile shows a symmetry about $R\sim  -0\farcs15$. The spike in $\sigma_{\mathrm{gas}}$ at $R = 0\farcs05$ disappears with rebinning and thus we ignore it. Finally, the fourth panel shows that the equivalent width of H$\alpha$ and [\ion{N}{2}] emission clearly peaks at $R \sim -0\farcs15$ and not $R =0\arcsec$.}

{In the \ion{Ca}{2} data, we find a stellar velocity dispersion peak of $132 \pm 19\ \kms$ in the bin closest to the gas kinematic center ($R\sim -0\farcs22$). This peak is larger than all other 7 bins, which range 90--100 \kms for the other 7 bins.  Consequently, we take this to be the kinematic center.}

\subsection{HET Spectra}

The STIS stellar absorption line data only extend to about 1\arcsec before the signal to noise becomes low enough to make the data unuseful, and we require additional spectroscopic data in order to measure the mass-to-light ratio and the stellar orbital structure, both of which are important for a measure of the black hole mass. The large radial spectroscopic data come from \citet{2015ApJS..218...10V}, using the Marcario Low Resolution Spectrometer \citep{1998SPIE.3355..375H} on the Hobby-Eberly Telescope (HET). {The HET spectra come from two slit positions: along the major axis and along the minor axis.}  The reductions follow \citet{2012Natur.491..729V} and \citet{2015ApJS..218...10V}. The HET spectra extend to 90\arcsec, and allow us to calculate the slit-averaged effective velocity dispersion $\sigma_e$ as
\begin{equation}
    \sigma_e^2 = \frac{\int_0^{R_e} (\sigma^2(r) + V^2(r))I(r) dr}{\int_0^{R_e} I(r) dr} = 104\ \units{km\ s^{-1}}.
    \label{eq:sigmae}
\end{equation}

\begin{figure*}[ht!]
\includegraphics[width=\textwidth]{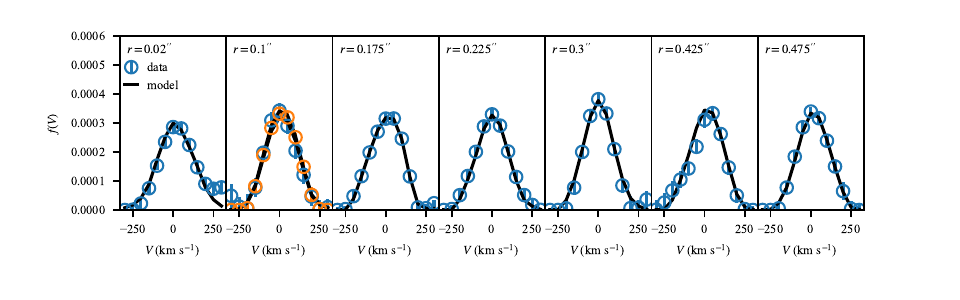}
\caption{LOSVDs from \emph{HST}/STIS observations.  Each panel shows the distribution of velocities in the major-axis spatial bin indicated by the radial coordinate from kinematic center.  For the $r = 0\farcs1$ bin, we plot the LOSVDs from both sides in the same panel, with the sign of velocity flipped so that they may be directly compared.  The error bars indicate the $1\sigma$ uncertainties.  Overall, the data show strong symmetry and high signal to noise, making these data good candidates for axisymmetric modeling. The best-fit model (black line) reproduces all of the major features.
\label{fig:hstlosvd}}
\end{figure*}

\begin{figure*}[ht!]
\includegraphics[trim=0 15 0 10 width=\textwidth]{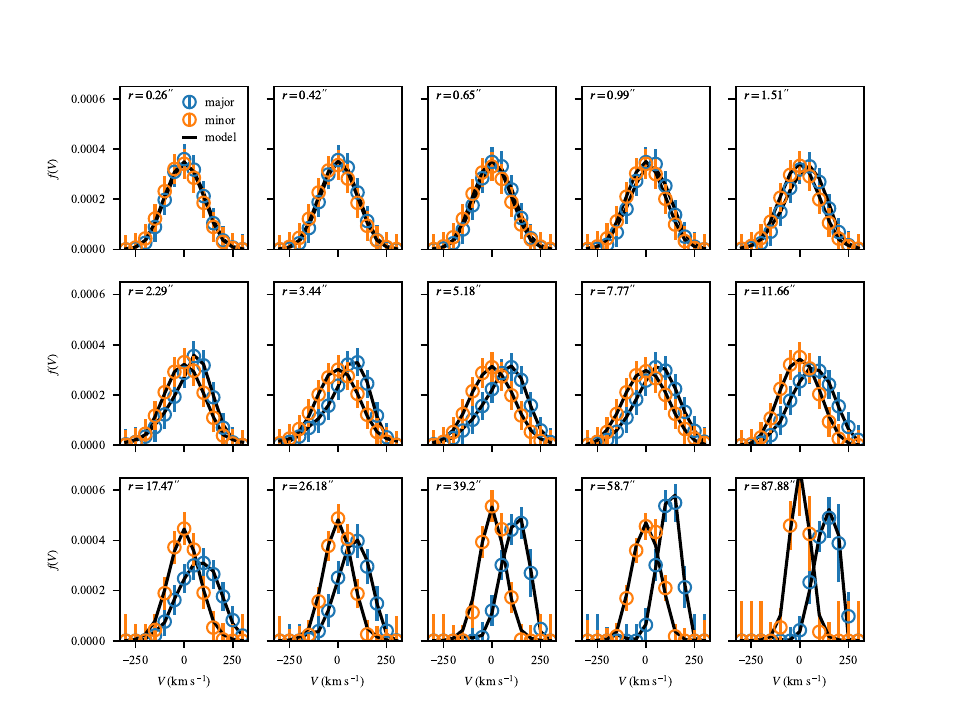}
\caption{LOSVDs from HET observations.  Each panel shows the distribution of velocities along the major (blue) and minor (orange) axes in the spatial bins indicated in each panel as distance from the kinematic center.  The major-axis LOSVDs show rapid rotation in the outer parts of the galaxy, where the light is dominated by the disk, whereas the minor-axis LOSVDs are centered at $V = 0$, as would be expected in an axisymmetric system.  The black line shows the best-fit model.
\label{fig:hetlosvd}}
\end{figure*}

From the one-dimensional spectrum we extracted line-of-sight velocity distributions (LOSVDs) as described in \citet{2009ApJ...695.1577G}, using the stellar template library due to \citet{2003ApJ...583...92G}.  We experimented with several different spatial binning schemes to find the best combination of spatial resolution and signal to noise ratio (S/N).  The best combination was a range of bin sizes from 0\farcs05--0\farcs55.
We show the extracted LOSVDs in Figs.\ \ref{fig:hstlosvd} and \ref{fig:hetlosvd} as well as the LOSVDs of the best-fit models from section \ref{mod}.  {In the second panel of Fig.\ \ref{fig:hstlosvd} ($r = 0\farcs1$) we have plotted LOSVDs from both sides of the galaxy, which are consistent with each other, giving us confidence that the stellar LOSVDs are not compromised by uneven dust absorption.}

\section{Modeling}
\label{mod}

\begin{figure}[ht!]
\includegraphics[width=\columnwidth]{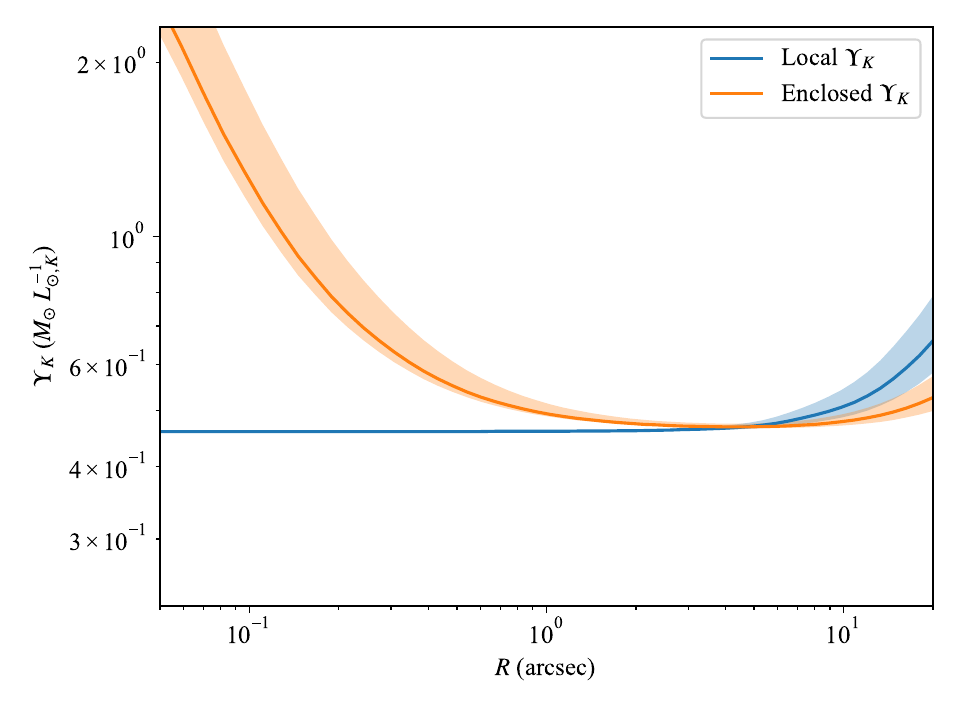}
\caption{Mass-to-light ratio radial profiles both local (solid blue curve) and enclosed (dashed orange curve).  These profiles come from the 20 models with the lowest $\chi^2$ and show the range of plausible values.  The local $\Upsilon_K(R)$ shows a flat inner portion with a slight increase starting at $R \approx 10\arcsec$.  The encolsed $\Upsilon(R)$ shows the black hole dominating the total mass-to-light ratio inside of $R = 1\arcsec$.
\label{fig:moverlrad}}
\end{figure}

We model NGC 4826 using the \citet{1979ApJ...232..236S} orbit library method implemented by the axisymmetric three-integral code described in \citet{2003ApJ...583...92G} and \citet{2009ApJ...693..946S}. 
Our modeling covers the classical bulge, pseudobulge components, and the region of the disk interior to radius 150\arcsec (5.3 kpc).  NGC 4826 is sufficiently close that we did not initially fit with a dark matter halo because it is impractical to get data out to large radii where the dark matter halo density is likely to have a significant gradient.  We found, however, that our modeling produced noticeably superior fits when including it.  This may be a result of what could be a small mass-to-light ratio gradient at the outer portions of our spectroscopic coverage, which is well modeled by a cored logarithmic halo model with small core radius ($r_c$).  We show the best-fit mass-to-light ratio as a function of radius in Fig.\ \ref{fig:moverlrad}.  There is a clear preference for an increase in $M/L$ outside of $R \approx 10\arcsec$, where the exponential disk begins to dominate the surface brightness profile (Figs.\  \ref{fig:photmodel} and \ref{fig:decomp1r}).  This is consistent with the photometric analysis of more recent star formation in the central pseudobulge than in the disk (section \ref{obs:imag:decomp}).
We used an iterative approach in the four-dimensional parameter space to find a global best fit and then ran a high-density grid around the best-fit location to best constrain the models.  In the end, our parameter space covers $M = 0$--$2 \times 10^{7}\ \msun$,  $\Upsilon_V = 0.3$--$5.25\ \msun\ L_{\scriptscriptstyle\odot}^{-1}$, $V_c = 90$--$190\ \kms$, and $r_c = 0.9$--$2.2\ \units{kpc}$.  Here 
$M$ is the mass of the black hole,  $\Upsilon_V$ is the $V$-band mass-to-light 
ratio, assumed to be the same for all five photometric components, and 
the gravitational potential of the spherical dark halo is assumed to be
\begin{equation}
    \Phi(r) = \frac{1}{2}V_c^2 \ln\left(r^2+r_c^2\right),
\end{equation}
where $r_c$ is the potential core radius and $V_c$ is the asymptotic velocity \citep{1996MNRAS.281...27P}.

The modeling results that we present are based on the assumption that the black hole is located at the \ion{Ca}{2} velocity dispersion peak of the galaxy rather than the surface brightness peak.  In most galaxies, the dispersion peak and the surface brightness peak are colocated at the center of the galaxy.  NGC 4826, however, has its dispersion maximum located $-0\farcs225$ from the surface brightness maximum.  As described above (\S \ref{obs:spec:halphaspec}), the most likely explanation for this is that obscuration from dust is slightly higher at the true center of the galaxy, which is intrinsically the brightest and has the highest velocity dispersion.  The fact that the gas emission line kinematics has a slightly different location for the dispersion profile center compared to the stellar absorption lines ($-0\farcs15$ compared to $-0\farcs225$, respectively) is not significant given the different binning used.  To determine the best location of the kinematic center and assumed location of the black hole, we modeled the entire system for seven locations along the STIS slit (one for each STIS spatial bin), finding that the two best were for the surface brightness maximum and velocity dispersion maximum.  
%% Dear arXiv comment reader: we ran so many models. The above sentences do not reflect the amount of work that went into determining what the heck was going on. 
We further refined the location by exploring sub-bin centering at the velocity dispersion maximum because of the substantial amount of rotation seen in this spatial bin, which an axisymmetric model cannot produce if the kinematic center is in the center of the pixel.  The best results were obtained when the center was assumed to be 0\farcs02 from the center of the 0\farcs05-wide (1 pixel) spatial bin.  Because of the goodness of fit and smoothness of the rotation curve at \emph{HST} resolution, we adopt this as our preferred answer.

\begin{figure*}[hb]
\centerline{\includegraphics[width=\textwidth]{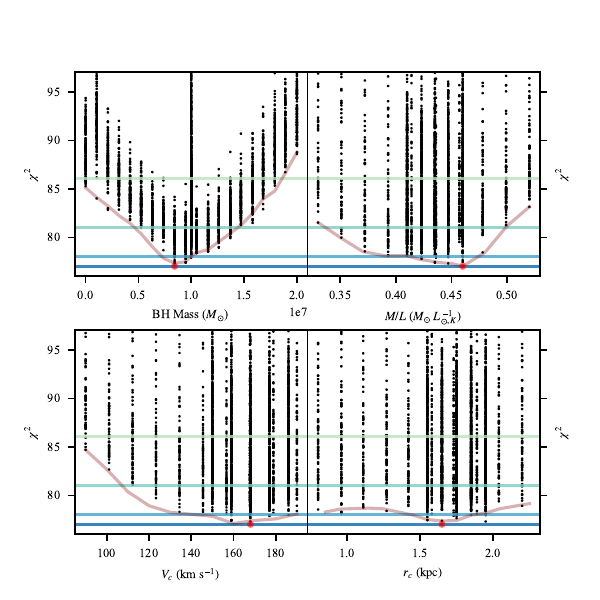}}
\caption{Modeling results.  Each panel plots $\chi^2$ against one of the model parameters (clockwise from top-left: black hole mass, mass to light ratio, dark matter halo core radius, and dark matter halo asymptotic velocity).  Each model is plotted with a black circle, and the best-fit model is plotted with a larger red circle.  The purple curves are smoothed traces of the minimum $\chi^2$ at each parameter value and show the marginalized $\chi^2$ curve.  From bottom to top, the horizontal lines show the minimum $\chi^2$, $\chi_{\mathrm{min}}^2 + 1$, $\chi_{\mathrm{min}}^2 + 4$, $\chi_{\mathrm{min}}^2 + 9$ to show 1, 2, and $3\sigma$ regions for one interesting parameter.  The $\chi^2$ curves are unimodal and {roughly} parabolic near the minimum, indicating robust results.  The best-fit mass with $1\sigma$ uncertainty is $M = 8.4^{+1.7}_{-0.6} \times 10^{6}\ \msun$, and is marginally consistent with $M = 0$ at about the $3\sigma$ level. \label{fig:chisq}}
\end{figure*}

We present the modeling results in Figure \ref{fig:chisq}.  Overall, the model is a good fit to the data.  
The results show smooth $\chi^2$ contours with roughly parabolic shapes in individual parameters.  Our best-fit parameters with $1\sigma$ uncertainties are $M = 8.4^{+1.7}_{-0.6} \times 10^{6}\ \msun$,  $\Upsilon_K = 0.46 \pm 0.03$, $V_{c} = 168 \pm 15\ \kms$, and $r_c = 1.65 \pm 0.24\ \units{kpc}$.  We calculate our uncertainties by taking the marginalized $\chi^2$ curve for each parameter and linearly interpolating to a $\Delta\chi^2 = n^2$ for an $n\sigma$ uncertainty.  {We tried alternative estimates of the best fit and uncertainties such as fitting a parabola to the smallest $\chi^2$ values for each value of $M$ with a variable noise parameter.  These alternative estimates resulted in similar uncertainty estimates and the minimum was usually in our $1\sigma$ interval and always in our $2\sigma$ interval.}  We are not able to rule out $M=0$ at $3\sigma$ as the best-fit $M = 0$ model is marginally consistent at $3\sigma$.  The core radius of the dark matter halo is poorly constrained with a large range of values able to produce acceptable fits.

\begin{figure}[htb]
\includegraphics[trim=0 10 0 10 , width=\columnwidth]{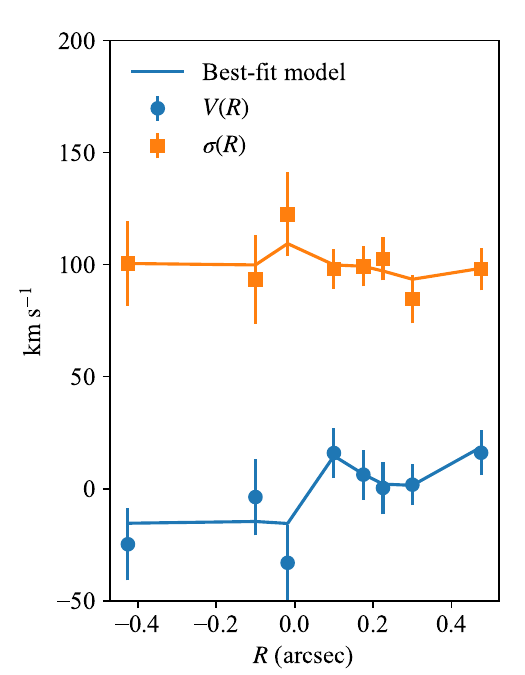}
\caption{Velocity and velocity dispersion profiles from \emph{HST}/STIS data and the best-fit model.  Although the model is not fit in this space, it is shown to produce a good match to the profiles and demonstrates how the dispersion peak is just off of $R = 0$ with substantial rotation about the center.
\label{fig:vsprofile}}
\end{figure}

\begin{figure}[htb]
\includegraphics[width=\columnwidth]{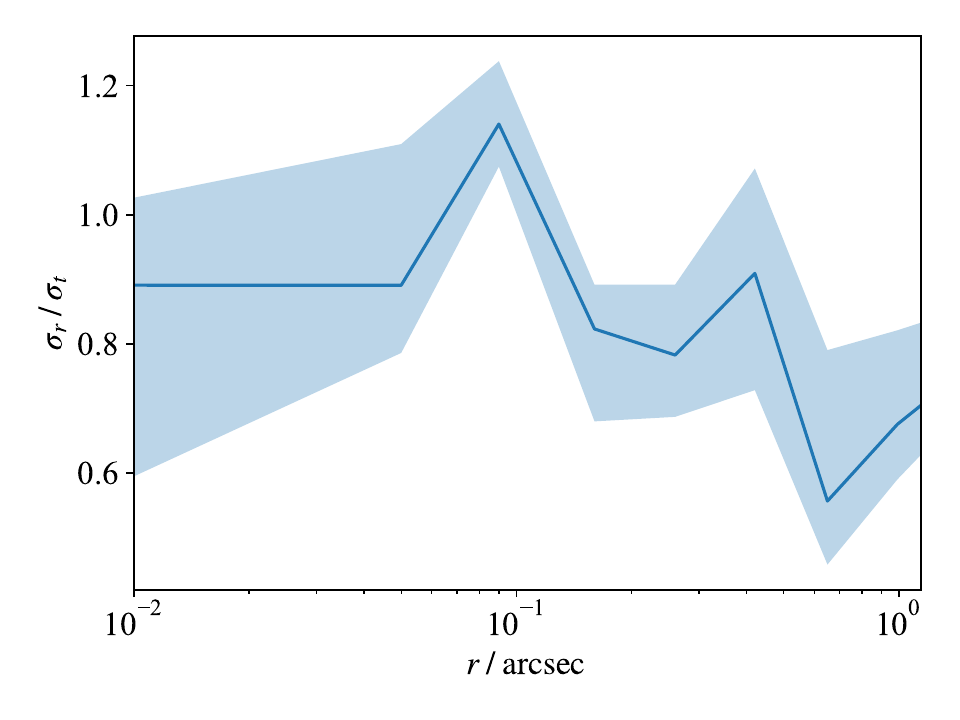}
\caption{Orbital anisotropy of the models.  We plot, as a function of radius, the ratio of radial velocity dispersion to the tangential (azimuthal and polar) velocity dispersion.  The solid line is from the best-fit model, and the shaded region comes from the range of values of the 20 best-fit models to estimate the uncertainty.
At large radii, the orbits are tangentially biased owing to the disk.  At smaller radii, the tangential bias may indicate a depletion of radial orbits from {the} earlier dynamical activity {that produced the counter-rotating gas streams}.
\label{fig:anis}}
\end{figure}

The velocity and velocity dispersion profiles from the best-fit models are plotted against the data in Fig.\ \ref{fig:vsprofile}.  The internal moments of the orbits are shown in Figure \ref{fig:anis}.  The anisotropy is shown as $\sigma_r / \sigma_t$, the ratio of the radial velocity dispersion of the stellar orbits ($\sigma_r$) to the tangential velocity dispersion (defined as the total rms tangential velocity, not merely relative to the mean rotational velocity; i.e., $\sigma_t = [0.5(\sigma_\phi^2 + \sigma_\theta^2)]^{0.5}$).  The anisotropy is tangentially biased almost everywhere.  At large radii ($r > 10\arcsec$), the tangential bias is a result of the stellar disk, which will obviously have primarily tangential motion.  Inside of $r = 0\farcs2$, the tangential bias may be a result of the dynamical event that produced the counter-rotating gas streams.  It is worth pointing out that we do not see any evidence of counter-rotating stellar motions.

\section{Discussion}
\label{discuss}
The evidence for a central dark mass from our modeling raises several issues. First is whether other data are consistent with a black hole existing in the galaxy nucleus.  Second are the implications of detecting such a low-mass black hole with Schwarzschild modeling.  Third is the matter of how such a low-mass black hole compares to predictions from scaling relations.  Fourth, the relevance to \emph{LISA} gravitational wave observations needs to be considered.  Last, we raise a question regarding the quenched nature of NGC4826 given its recent merger.  We discuss each of these in turn below.  
\label{discussion}
\label{ponderings}

\subsection{A low-mass black hole measured with Schwarzschild modeling}
\label{discuss:smallest}

NGC 4826 hosts one of the smallest black holes measured with the stellar dynamical Schwarzschild modeling technique.  At $8.4 \times 10^{6}\ \msun$, NGC 4826's black hole is within a factor of 4 of the black hole in NGC 221 (M32), which has mass $M = (2.5 \pm 1.0) \times 10^{6}\ \msun$ \citep{2010MNRAS.401.1770V}, the  smallest mass determined by Schwarzschild modeling.  There are Schwarzschild modeling results with upper limits below the mass of NGC 221 such as M33 ($M < 1.5 \times 10^{3}\ \msun$ in \citealt{2001AJ....122.2469G} and $M < 3 \times 10^{3}\ \msun$ in \citealt{2001Sci...293.1116M}) and NGC 205 \citep[$M < 3.8 \times 10^{4}\ \msun$;][]{2005ApJ...628..137V}.  In both of these cases, there is no external evidence for the presence of a black hole so that both of these dynamical results can be interpreted as evidence for absence of a black hole, though \citet{2019ApJ...872..104N} find dynamical evidence for the presence of a black hole in NGC 205 of mass $M = 7_{-7}^{+96} \times 10^{3}\ \msun$.

There are smaller known black holes using other mass estimation techniques both direct and indirect, including those in NGC 4395 ($M = 4.0^{+8}_{-3} \times 10^{5}\ \msun$ from ionized gas dynamics due to \citealt{2015ApJ...809..101D}; $10^{4}$--$10^{5}\ \msun$ from reverberation mapping due to \citealt{2005ApJ...632..799P, 2019NatAs...3..755W}), NGC 404 ($M = 5.5^{+4.1}_{-3.8}\times 10^5\ \msun$ from molecular gas dynamics due to \citealt{2020MNRAS.496.4061D}), Pox 52 ($M \approx 2$--$4\times10^{5}\ \msun$ from a variety of indirect methods due to \citealt{2004ApJ...607...90B, 2008ApJ...686..892T}), and SDSS J152303.80+114546.0 (RGG 118, $M\sim5\times10^{4}\ \msun$ based on broad emission line measurements due to \citealt{2015ApJ...809L..14B}).

Schwarzschild modeling in particular and dynamical modeling in general is sensitive to low-mass black holes, and dynamical mass measurements of black holes are not biased high as has been claimed in the literature.  
While it is true that\,---\,all else equal\,---\,smaller black holes are more difficult to detect than larger black holes, the claims that this leads to individual mass estimates and population estimates to be biased to high masses \citep[e.g.,][]{2016MNRAS.460.3119S} are not valid.  The fact that our modeling sometimes yields upper limits consistent with zero shows that black holes with spheres of influence too small to be resolved will not always result in a false detection of a mass with an implied resolvable sphere of influence \citep{2004astro.ph..3257R, 2020IAUS..353..186K}.  While there is the possibility of a publication bias in which upper limits will not be written about, previous analysis has shown that the mass scaling relations and their intrinsic scatter measurements are unlikely to be biased \citep{2011ApJ...738...17G, 2020IAUS..353..186K}.

\subsection{Comparison to empirical black hole mass scaling relations}
\label{discuss:scaling}

Here we show that the black hole mass $M = 8.4^{+1.7}_{-0.6} \times 10^{6}\ \msun$, or $\log{M} = 6.92^{+0.08}_{-0.03}$ in the composite (classical bulge plus pseudobulge) galaxy NGC 4826 is consistent
with the black hole--host galaxy correlations separately derived for galaxies that have dominant classical bulges and those that have dominant pseudobulges.  Our comparisons are made using bulge properties derived here in Section 2 and the Appendix and the $K_s$-band mass-to-light ratio $M/L = 0.46$ derived above.  The velocity dispersion $\sigma_e = 104\ \kms$ is calculated from our data as described in section \ref{obs:spec} and Eq.\ (\ref{eq:sigmae}).

\begin{figure*}[htb]
\includegraphics[width=1.05\textwidth]{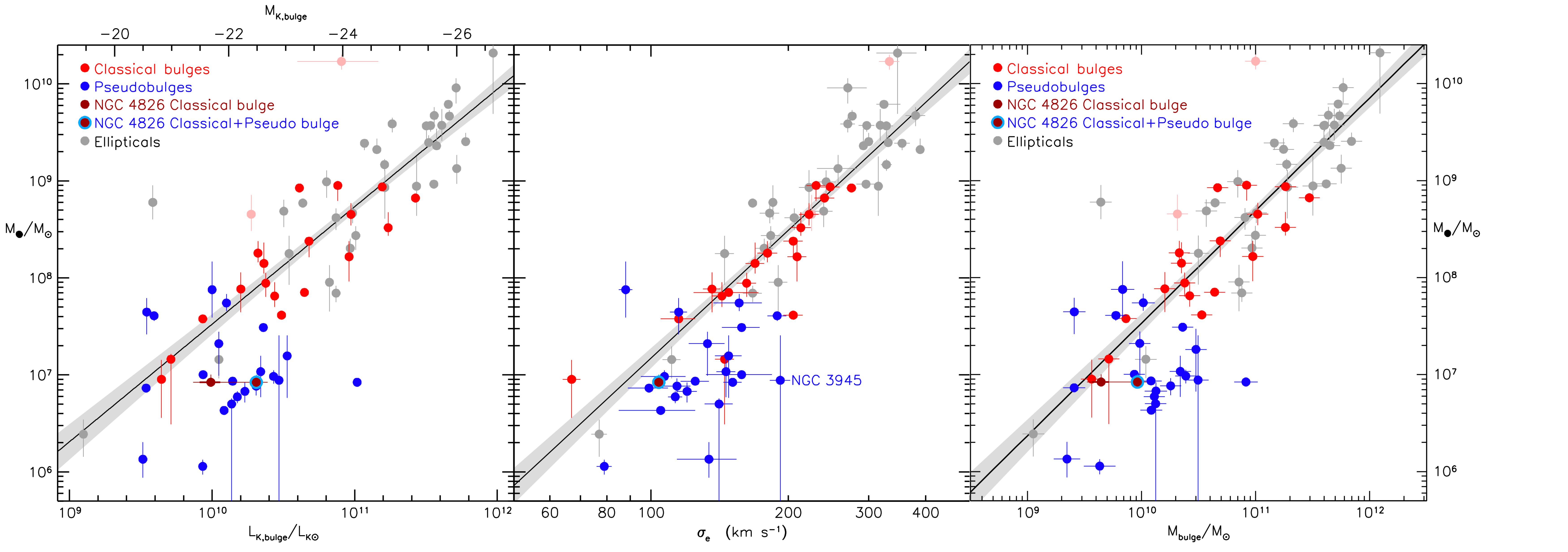}
%\vspace{-170pt}
\caption{Correlations of black hole mass (here designated M$_\bullet$) with ({left}) the $K_s$-band absolute magnitude and luminosity of the bulge component of the galaxy, ({center}) the bulge velocity dispersion, and ({right}) the stellar mass of the bulge component.  These are the correlations from \citet{2013ARA&A..51..511K} with the preliminary black hole mass for NGC 4826 replaced with the presently derived value and with the point plotted twice, once (dark red) for the adopted classical bulge parameters and again for the sum of the classical and pseudo bulge (dark red point surrounded by light blue).  The thick horizontal error bars on the classical bulge points are for the $1\sigma$ uncertainty in $B/T$ quoted in Figures 5 and 6.  The thin horizontal error bars end at the bulge parameters for the extreme ``error bar" decompositions illustrated in the Appendix.
\label{fig:correlations}}
\end{figure*}

Figure \ref{fig:correlations}  shows the black hole correlations from \citet{2013ARA&A..51..511K}.  We use these
for our main comparison (1) because they omit black holes whose masses are underestimated because the
widths of spectral emission lines were not taken into account in dynamical analyses, (2) because mergers in
progress are shown in the above paper to not participate in the correlations and therefore are omitted, (3) because masses measured via stellar dynamics included dark matter halos in most cases, and (4) because the distinction between classical and pseudo bulges was measured in the most detail available.  So the derived relations were as robust as 2013 data allowed.  The scatter in the relations for classical bulges and ellipticals was correspondingly small and essentially the same for all three correlations, $0.29 \pm 0.01$ dex.  {Other empirical measurements of black hole mass correlating with host galaxy properties exist  \citep[e.g.,][]{
1993nag..conf..197K,  1995ARA&A..33..581K,  1998AJ....115.2285M,  2000ApJ...539L...9F,  2000ApJ...539L..13G,  2001ApJ...563L..11G,  2002ApJ...574..740T,  2003ApJ...589L..21M,  2004ApJ...604L..89H,  2007ApJ...669...67H,  2008ApJ...680..143G,  2009ApJ...698..198G,  2010ApJ...716..269W,  2010ApJ...720..516B,  2011ApJ...739...28X,  2012MNRAS.419.2497B,  2013ApJ...764..184M,  2013ApJ...778...47S,  2015ApJ...813...82R,  2016ApJ...826L..32G,  2016ApJ...818...47S,  2016ApJ...831..134V,  2017MNRAS.471.2187D,  2018ApJ...869..113D,  2019ApJ...871...80G,  2019ApJ...873...85D,  2019ApJ...876..155S,  2019ApJ...887...10S,  2019MNRAS.490..600D,  2020IAUS..353..186K,  2021ApJ...921...36B,  2022ApJ...927...67S,  2023ApJ...956...60C,  2023MNRAS.520.1975G, 2024MNRAS.531..230G}.  We use the \citet{2013ARA&A..51..511K} relation instead of any of the others because of the four points mentioned above.  The other measurements are not as useful for furthering understanding about the physics underlying the scaling relations because of one or more of the following: (1) they do not have the same quantity of quality data, (2) they are intended for black hole mass estimation at large distances where even bulge--disk decomposition is difficult let alone classical--pseudobulge decomposition, or (3) they are investigating different physics such as AGN-related scaling relations.}

Figure \ref{fig:correlations} shows that the black hole in NGC 4826 satisfies all three correlations if we use the parameters of the classical bulge part of the galaxy.  In the $M$--bulge luminosity correlation, the classical bulge point lies low but is consistent with the range of the scatter.  In the other two correlations, the classical bulge point lies close to the ridge line of the scatter.  

The blue points in Figure \ref{fig:correlations} are for galaxies which have only a pseudobulge or which were judged to have a pseudobulge that is substantially more massive than any classical bulge.  NGC 4826 was recognized to be a difficult case with comparable classical and pseudo bulges.  This is consistent with the conclusions of the present paper.  Lacking the detailed measurements and analysis of the present work, the galaxy was plotted then as a blue point with a preliminary black hole mass measurement.  This has been replaced here with the present data.  If we treat NGC 4826 in the same way as the blue points, i.e.,  if we plot a point at the sum of the classical bulge and pseudobulge luminosity of mass, then that point is low---outside the scatter for the red points---in the correlation with $K_s$-band luminosity.  This point is consistent with the conclusions of 
\citet{2008MNRAS.386.2242H}, \citet{2010ApJ...721...26G}, \citet{2011Natur.469..374K}, and 
\citet{2013ARA&A..51..511K}
that pseudobulges contain undermassive black holes---undermassive when compared to the correlation for classical bulges and ellipticals.  That is, the blue-plus-red point for NGC 4826 is essentially in the middle of the (uncorrelated) scatter in the blue points.  

In the $M$--$\sigma$ correlation, black holes in pseudobulges also tend to be undermassive compared to those in classical bulges, but their mean deviation below the classical bulge relation is less than it is for the bulge mass and luminosity correltions.  The black hole in NGC 4826 is almost on the ridge line of the correlation for classical bulges and ellipticals, as is the case for many other pseudobulges.

Note in the left-hand parts of Fig.\ \ref{fig:correlations} that there is very little ``room" for the process that made the pseudobulge to have added significantly to the black hole mass.  This is consistent with the suggestion in \citet{2013ARA&A..51..511K} that black hole growth has little effect on galaxy evolution for such galaxies.

We can also compare to other published correlations.
In contrast to the bulge scaling relations presented above, maser measurements show a broader distribution of masses at fixed galaxy property.  Using a 
sample that includes 20 black holes with maser mass estimates, \citet{2016ApJ...826L..32G} found an $M$--$\sigma_{*}$ relation for late-type galaxies. The definition of velocity dispersion used by \citet{2016ApJ...826L..32G}, i.e., $\sigma_{*}$,  does not include rotational support so that the corresponding value for NGC 4286 ($\sigma_{*} = 92\ \kms$) is smaller than our earlier value of $\sigma_e$.  The predicted mass is $\log{M} = 6.65 \pm 0.49$, consistent with our mass estimate.   It is worth noting that among dynamical mass measurements, the $M$--$\sigma$ and $M$--$L_{\mathrm{bulge}}$ relations are poorly populated near NGC 4826's location.  In particular, only two galaxies have smaller velocity dispersions than NGC 4826 in the \citet{2013ARA&A..51..511K} sample and only six in the \citet{2016ApJ...826L..32G} sample, so that the $M$--$\sigma$ relation is necessarily less well measured here.

We also compare to the \citet{2015ApJ...813...82R} relation between the black hole mass of an AGN and the total stellar mass of the host, regardless of the bulge type.  Adopting a solar $K$-band absolute magnitude of $K\subsun = 3.27$ \citep{2018ApJS..236...47W}, our measured total $K$-band absolute magnitude of $M_{KT} = -24.06$ for NGC 4826 implies a $K$-band stellar luminosity $\log{(L_T / L\subsun)} = 10.93$.  Using our estimated $M/L = 0.46$, the total stellar mass is $\log{(M_{*} / \msun)} = 10.59$.  This stellar mass predicts a logarithmic black hole mass mass of $\log{M} = 7.02 \pm 0.55$, very close to our measured mass.

\subsection{Multiwavelength evidence for a black hole in NGC 4826}
\label{discuss:multiwavlength}
Our black hole mass estimate is only marginally consistent with $M = 0$ at the $3\sigma$ level, and multiwavelength data  support the presence of a mildly accreting black hole.  The most compelling indication for an accreting black hole in NGC 4826 comes from optical spectra due to \citet{2010ApJS..190..233M}.  The spectra were taken in a square nuclear region of size 2\farcs5 as well as at larger angular scales.  Although the nuclear spectrum shows no broad Balmer lines, the narrow line ratios indicate the power source of ionizing radiation to be an AGN.  The values and uncertainties in [\ion{N}{2}] / H$\alpha$ and [\ion{O}{3}] / H$\beta$ allow for the possibility that the lines are driven by a composite AGN and starforming ionizing radiation field, but it is nominally in the  \citet{2001ApJ...556..121K} AGN region of the \citet[][BPT]{1981PASP...93....5B} diagram and in any case still requires some contribution from an AGN.  A potential complication is that at low luminosities and perhaps especially so for low-mass black holes, it is difficult to identify shock-ionization, which may result from either AGN or from stellar sources \citep{2008ARA&A..46..475H, 2010ApJ...711..796E, 2018ApJ...864...90M}.  {Our analysis of the higher spatial resolution STIS emission-line spectrum is mostly in agreement with the results from \citet{2010ApJS..190..233M} with even higher [\ion{N}{2}] / H$\alpha$ values (1.45--2.6 depending on the spatial location of the emission, compared to their value of 1.2--1.3) thus increasing the confidence of optical-emission-line identification as an AGN.  We finally note that NGC 4826 was also classified as an AGN in \citet{1997ApJS..112..315H}.}

\begin{figure*}[htb]
\includegraphics[width=\textwidth]{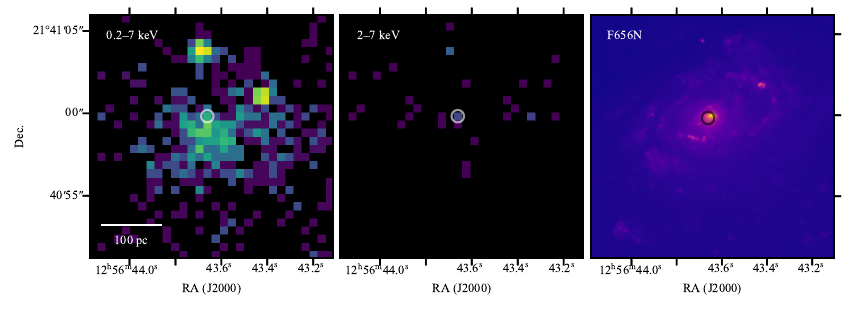}
\caption{Central $15 \times 15\arcsec$ of NGC 4826 as observed with \emph{Chandra} in the 0.2--7 keV band (left), the hard 2--7 keV band (center), and with \emph{HST}/WFPC2 through the F656N filter.  The full band X-ray image shows distinct point sources as well as diffuse emission.  The hard band X-ray image shows a relatively small amount of emission coming from central region, but there is a potential hard X-ray point source (location indicated by a circle in all three panels) that is consistent with the stellar surface brightness peak, though it requires deeper X-ray imaging to fully assess.
\label{fig:xray}}
\end{figure*}

The X-ray data of NGC 4826 do not strongly argue for an accreting black hole, nor do they present strong evidence against the case.  As part of a larger \emph{Chandra} archival X-ray survey of the \emph{Spitzer} Infrared Nearby Galaxy Survey, \citet{2011ApJ...731...60G} note that NGC 4826 has a nuclear X-ray point source but that it is unlikely to be AGN emission because of its soft spectrum in a 2 ksec observation.  Since then, a deeper 30 ksec observation was taken and has largely confirmed the picture that the central X-ray emission is dominated by a combination of dimmer X-ray point sources (presumably X-ray binaries) and diffuse emission.  We show an image of the \emph{Chandra} data in Figure \ref{fig:xray}.  

Our analysis of the newer \emph{Chandra} data, which were also analyzed by \citet{2019ApJS..243....3L} as part of a study on the relation between X-ray binary luminosity functions and host galaxy star formation rates, leads us to conclude that the central X-ray emission is dominated by extended, soft X-ray emission at the center with the potential for a hard (2--7 keV) point source at the putative location of the nuclear black hole but with too few counts to be decisive.  

We used the \citet{2019ApJ...877...17F} Bayesian AnalYsis of Multiple AGN in X-ray (BAYMAX) code to test for the presence of a point source.  We compared two models for the X-ray emission in the central $5\times5\arcsec$ region: (i) a model of uniform, diffuse X-ray emission and (ii) a model with uniform, diffuse X-ray emission plus a point source at some location in the central region.  Our analysis strongly prefers the model with a point source, the best-fit location of which has large uncertainties but is consistent with any definition of the optical/infrared center of the galaxy.  We note that our ability to infer the presence of a point source with BAYMAX depends on how well our diffuse emission model matches reality.  In this case, it would require a deeper observation to attempt with more sophisticated diffuse emission models.  Even if we take our analysis at face value, the presence of an X-ray point source does not necessarily imply the presence of an AGN.  The point source could be an X-ray binary (XRB) or a compact knot of gas heated by shocks or other external means \citep[e.g.,][]{2020ApJ...892...29F}.  Because the uncertainties on the location of the X-ray point source are also comparable to the separation between the dispersion peak and optical surface brightness peak, we do not attempt absolute astrometric calibration between the X-ray and optical imaging.    We also analyzed the X-ray spectra of the decomposed diffuse and point-source emission as absorbed power-laws and find that the point source is best described as a soft ($\Gamma = 6.1^{+2.4}_{-2.2}$, $3\sigma$ uncertainties) power-law with modest intrinsic absorption ($N_H = 5.1^{+13.2}_{-3.8}\times 10^{21}\ \units{cm^{-2}}$, $3\sigma$ uncertainties).  The luminosity of the point source in the 0.5--8 keV band is $L_X = 5.2^{+1.2}_{-2.4}\times10^{36}$ erg s$^{-1}$ ($3\sigma$ uncertainties).  The point source is notably harder than the best-fit diffuse emission model ($\Gamma = 7.6^{+1.6}_{-2.0}$, $3\sigma$ uncertainties), but is still softer than would be expected for AGN emission and has a luminosity that is low enough to be an XRB.

In general, it is possible to distinguish low-luminosity X-ray emission in an AGN from X-ray binary emission by using the fundamental plane of black hole accretion \citep{2019ApJ...871...80G}, but in the case of NGC 4826, the source of radio emission appears to be stellar.  The fundamental plane of black hole accretion relates the mass of the accreting black hole, its X-ray luminosity, and its core radio luminosity.  Because the relation is valid over a large range of black hole masses, it can be used to determine whether a given accreting black hole with X-ray and radio emission is an X-ray binary, and intermediate-mass black hole, or a supermassive black hole.  NGC 4826 does have a nuclear radio emission measurement from Karl G.\ Jansky Very Large Array observations at 1.4 GHz of $S_\nu = 103.4\ \units{mJy}$ \citep{1998AJ....115.1693C}.  Based on the ratio of far infrared fluxes to radio fluxes, \citet{2002AJ....124..675C} concluded that the radio emission was powered by star formation, compromising the utility of the use of the fundamental plane as a mass estimation tool.

Overall, the multiwavelength evidence is mildly supportive of our conclusion that there is a SMBH at the center of NGC 4826, which we measure to be in the range $M = 8.4^{+1.7}_{-0.6} \times 10^{6}\ \msun$ with 68\% confidence.

\subsection{Relevance to \emph{LISA}}
\label{discuss:lisa}

The relatively small mass of the black hole in NGC 4826 combined with evidence of recent merger activity makes it a prototype for the kinds of sources to which \emph{LISA} will be sensitive.  
The gravitational wave frequency of a merging black hole system right before coalescence scales as $M^{-1}$.  So \emph{LISA}'s good sensitivity in the range of $\sim2\times 10^{-3}$ to $2 \times 10^{-2}\ \units{Hz}$ corresponds to a peak mass sensitivity of $\sim10^{6}(1 + z)^{-1}\ \msun$.  Thus the existence of black holes with mass $\sim10^{5}$--$10^{7}\ \msun$ at $z=0$ indicates the potential for high signal-to-noise gravitational wave events with \emph{LISA}.

In addition to the mere existence of a low-mass black hole, the evidence of merger activity in NGC 4826 demonstrates such black holes may participate in mergers with other black holes.  The primary evidence of NGC 4826's interaction is its counter-rotating outer gas disk \citep{1992Natur.360..442B} and the presence of inner, leading spiral-arm structure and outer, trailing spiral-arm structure \citep{1994AJ....107..184W, 1994AJ....107..173R}.  This activity is present despite the isolated environment of NGC 4826.  The details of the gas disks likely indicate a low-mass merger and do not resolve the question of whether what has merged with NGC 4826 had a black hole, but it nonetheless increases confidence in the potential for low-mass black holes to merge with others in the local Universe.  Confidence in low-mass galaxies to merge is further boosted by the recent analysis of halo stars that show a shell feature and a plume feature that is well interpreted as remnants of a recent merger with a gas-rich galaxy of mass similar to NGC 292 \citep{2023ApJ...949L..37S}.

The evidence of recent merger activity makes NGC 4826 the most recent member of the ``mergers in progress'' with black hole mass measurement.  ``Mergers in progress'' tend to have undermassive black holes compared to expectations based on galaxy scaling relations \citep{2013ARA&A..51..511K}.  The mass we measure for NGC 4826 is smaller than any others identified in \citet{2013ARA&A..51..511K} and the only one consistent with $M < 1 \times 10^{7}\ \msun$ at the $1\sigma$ level, though two are just above (NGC 2960: $M = 1.08^{+0.04}_{-0.05} \times 10^{7}\ \msun$; IC 1381: $M = 1.49^{+0.44}_{-0.45} \times 10^{7}\ \msun$).

\subsection{Open Questions}
A remaining question is why, overall, star formation in NGC 4826 is quenched when it has manifestly
accreted a gas-rich companion recently enough so that we still see large amounts of dust and, at large radii, a counter-rotating gas disk.  Feedback from AGN activity is unlikely to be the cause of halting star formation.  
In the context of our $M$ measurement and the black hole correlations, there is little room for any significant addition to the 
black hole mass via the galaxy accretion event that{\thinspace}---{\thinspace}we argue{\thinspace}---{\thinspace}built the pseudobulge.   

If we underestimate the black hole mass by a factor of $\sim2$, then there is more room for recent black hole growth and therefore radiative output by an AGN. We offer a few caveats that are, effectively, uncontrollable systematic uncertainties that could conceivably contribute to a an underestimated mass.  The slight difference in location between the velocity dispersion and surface
brightness maxima could be a sign that near-central dust absorption is bigger than we think.  We therefore may not see the high-velocity stars 
that are signatures of a bigger black hole.  Additionally, STIS slit centering was carried out at $\sim 8500$ \AA.  In a worst-case 
scenario, asymmetric dust absorption could result in a slit centering that missed the black hole and thus we only see a diminished effect of a larger black hole.   Uncertainty about $M$ may be resolved with ALMA measurements, if there is a molecular gas disk near the center.

\section{Summary}
\label{sum}

We used high-spatial-resolution observations of NGC 4826 to analyze the light distribution and structural parameters of the host as well as to measure the black hole mass and mass-to-light ratio.  The imaging data were \emph{HST}/WFPC2 in F547M and F656N, 2MASS, NICMOS, WISE, and \emph{Spitzer}. The spectroscopic data of the nucleus were \emph{HST}/STIS observations taken of the \ion{Ca}{2} triplet and a series of emission lines, combined with HET LRS data of the outer regions of the galaxy.  
The photometric analysis and decomposition revealed complex morphology with a combination classical bulge and pseudobulge as well as a disk.  Our adopted decomposition requires an $n = 3.27$ classical bulge, an $n = 0.58$ partial pseudobulge, two lens components (an $n = 0.18$ inner lens and an $n = 0.17$ outer lens), and a disk component with $n = 1.07$.  We consider the partial pseudobulge and two lens components as the galaxy's pseudobulge.   The total $V$-band luminosity is $M_{VT} = -21.07$, the classical bulge to total ratio is $B/T = 0.12$, the pseudobulge to total ratio is $PB / T = 0.13$, and the disk to total ratio is $D/T = 0.75$.

Our Schwarzschild modeling of the galaxy found the mass-to-light ratio in the $K_s$ band to be $\Upsilon_K = 0.46 \pm 0.03\ \units{\msun\ L_{\scriptscriptstyle\odot}^{-1}}$ and the black hole mass to be $M = 8.4^{+1.7}_{-0.6} \times 10^{6}\ \msun$ at our assumed distance of 7.27 Mpc.  
 
Our modeling is only marginally consistent with $M = 0$ at about the $3\sigma$ level, and a preponderance of other lines of evidence, including BPT diagnostics of optical emission lines and continuum X-ray emission, leads us to conclude that there is an accreting black hole at the center of NGC 4826.  

This black hole mass in NGC 4826 satisfies published scaling relations between $M$ and the $K$-band luminosity, stellar mass, and
velocity dispersion of only the classical bulge part of the galaxy.  It is undermassive compared to the sum of the bulge and pseudobulge
$K$-band luminosities.  Thus the black hole in NGC 4826 is consistent with published results on scaling relations between black holes
and classical and pseudo bulges.  

The black hole in NGC 4826 is among the smallest to have been measured via orbit superposition modeling of the stellar dynamics.   The measurement of such a small black hole mass demonstrates the ability of black-hole-mass measurements to probe small masses.  
The presence of a black hole with mass $M\la10^{7}\ \msun$ in a galaxy with evidence for a recent merger is also promising as an example of low-mass black holes that may merge with other black holes, to which \emph{LISA} will be sensitive. \vskip 10pt

%% The ability to measure NGC 4826's black hole mass is insignificant next to the power of the Force.

\software{FORTRAN77, numpy, matplotlib, sm, Lick VISTA, NOAO IRAF}

\facilities{FLWO:2MASS, HET (LRS), HST (NICMOS, STIS, WFPC2), Spitzer, TACC, WISE}

\acknowledgments
We thank STScI's Hubble Heritage project for making their images available in the public domain.
KG{\"u} thanks Jessie Runnoe for helpful conversations regarding the AGN emission line spectra.
XVISTA is maintained and distributed by Jon Holtzman at
New Mexico State University ({\tt http://ganymede.nmsu.edu/holtz/xvista}).
Some of the data presented in this article were obtained from the Mikulski Archive for Space Telescopes (MAST) at the Space Telescope Science Institute. The specific observations analyzed can be accessed via \dataset[10.17909/yep1-qr33]{https://doi.org/10.17909/yep1-qr33}.  We have also made our photometry and LOSVD data available at \dataset[10.7302/kr8z-fj98]{https://doi.org/10.7302/kr8z-fj98}.

The authors acknowledge the Texas Advanced Computing Center (TACC) at The University of Texas at Austin for providing computational resources that have contributed to the research results reported within this paper.

This work made use of the following software packages: \texttt{FORTRAN77}, \texttt{Jupyter} \citep{2007CSE.....9c..21P, kluyver2016jupyter}, \texttt{matplotlib} \citep{Hunter:2007}, \texttt{numpy} \citep{numpy}, \texttt{python} \citep{python}, \texttt{SM}, \texttt{VISTA} \citep{1988igbo.conf..443S}, and \texttt{IRAF } \citep{1993ASPC...52..173T, 1986SPIE..627..733T}.  This research has made use of NASA's Astrophysics Data System.  Software citation information aggregated using \texttt{\href{https://www.tomwagg.com/software-citation-station/}{The Software Citation Station}} \citep{software-citation-station-paper, software-citation-station-zenodo}.

{JK is grateful for the long-term support provided by the
Curtis T. Vaughan, Jr. Centennial Chair in Astronomy.  We are
most sincerely grateful to Mr. and Mrs. Curtis T. Vaughan, Jr.
for their many-year support of Texas astronomy.
The ground-based spectrum of NGC 4826 was obtained with the 
Marcario LRS and the Hobby–Eberly Telescope (HET). LRS is named
for Mike Marcario of High Lonesome Optics; he made optics
for the instrument but died before its completion. LRS is a
project of the HET partnership and the Instituto de Astronom{\'i}a
de la Universidad Nacional Aut\'onoma de M\'exico. The HET is
a project of the University of Texas at Austin, Pennsylvania
State University, Stanford University, 
Ludwig-Maximilians-Universit\"at M\"unchen, and 
Georg-August-Universit\"at G\"ottingen.
The HET is named in honor of its principal benefactors, William
P. Hobby and Robert E. Eberly.}

\begin{figure}[htb]
\centering
\includegraphics[trim=15 10 25 30 , width=\columnwidth]{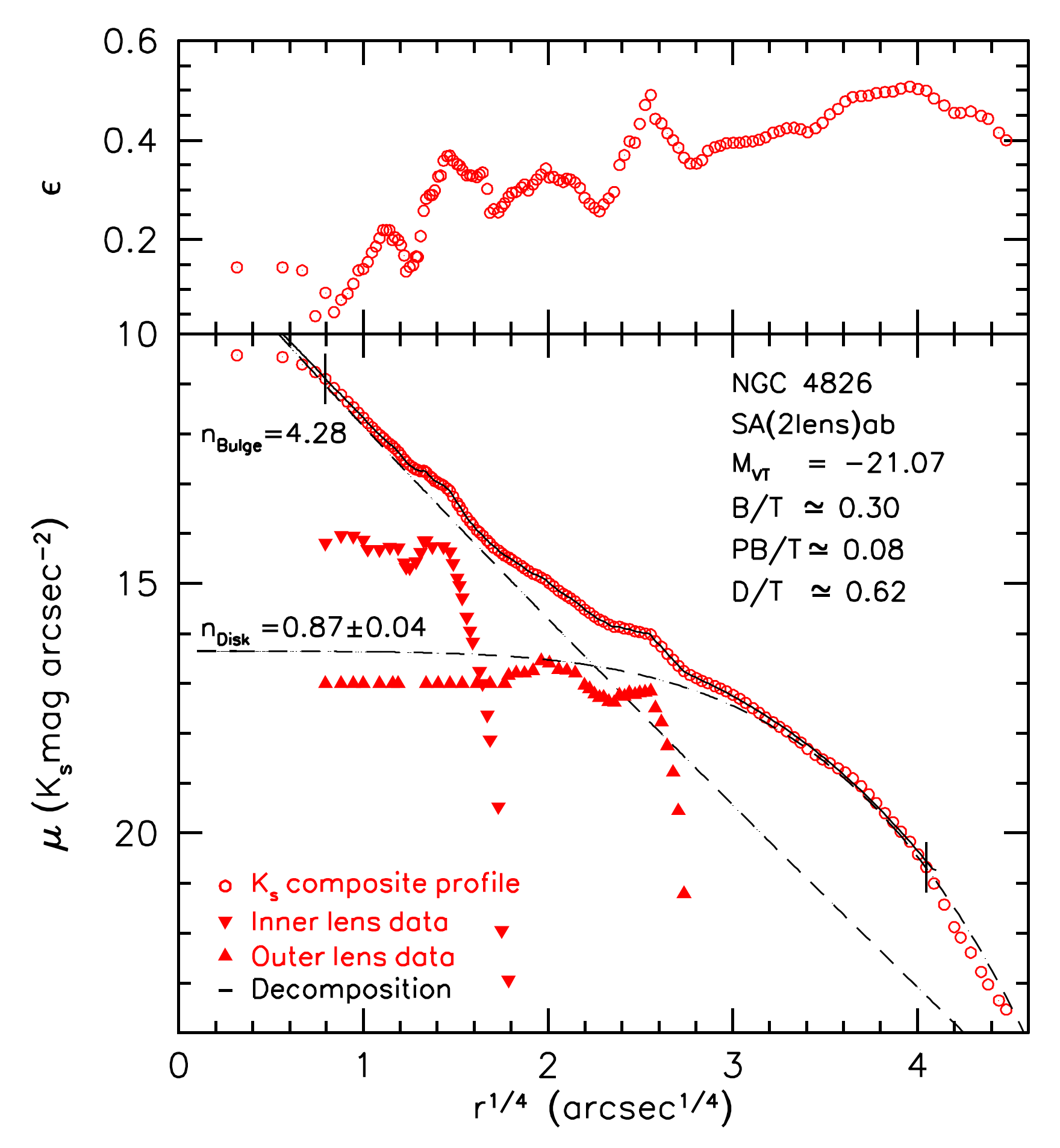}
\caption{
 Minimal-pseudobulge photometric decomposition plotted against $(r/\mathrm{arcsec})^{1/4}$ to show components 
 at both small and large radii. Here, the lens components are treated as they are in our adopted fit, and
 together, they are assumed to be all of the pseudobulge.  Thus all of the rest of the profile is decomposed 
 into two S{\'{e}}rsic components to represent a classical bulge and disk.  The statistical uncertainties in 
 this fit are small. However, the classical bulge has $n = 4.28$, larger than values normally seen in small 
 bulges, and the disk S{\'e}rsic index $n = 0.87 \pm 0.04 < 1$ is smaller than $n = 1$ for an exponential.  
 Data and symbols are the same as in Fig.\ \ref{fig:decomp1r4}.
\label{fig:decomp2r4}}
\end{figure}

\begin{figure}[htb]
\centering
\includegraphics[width=\columnwidth]{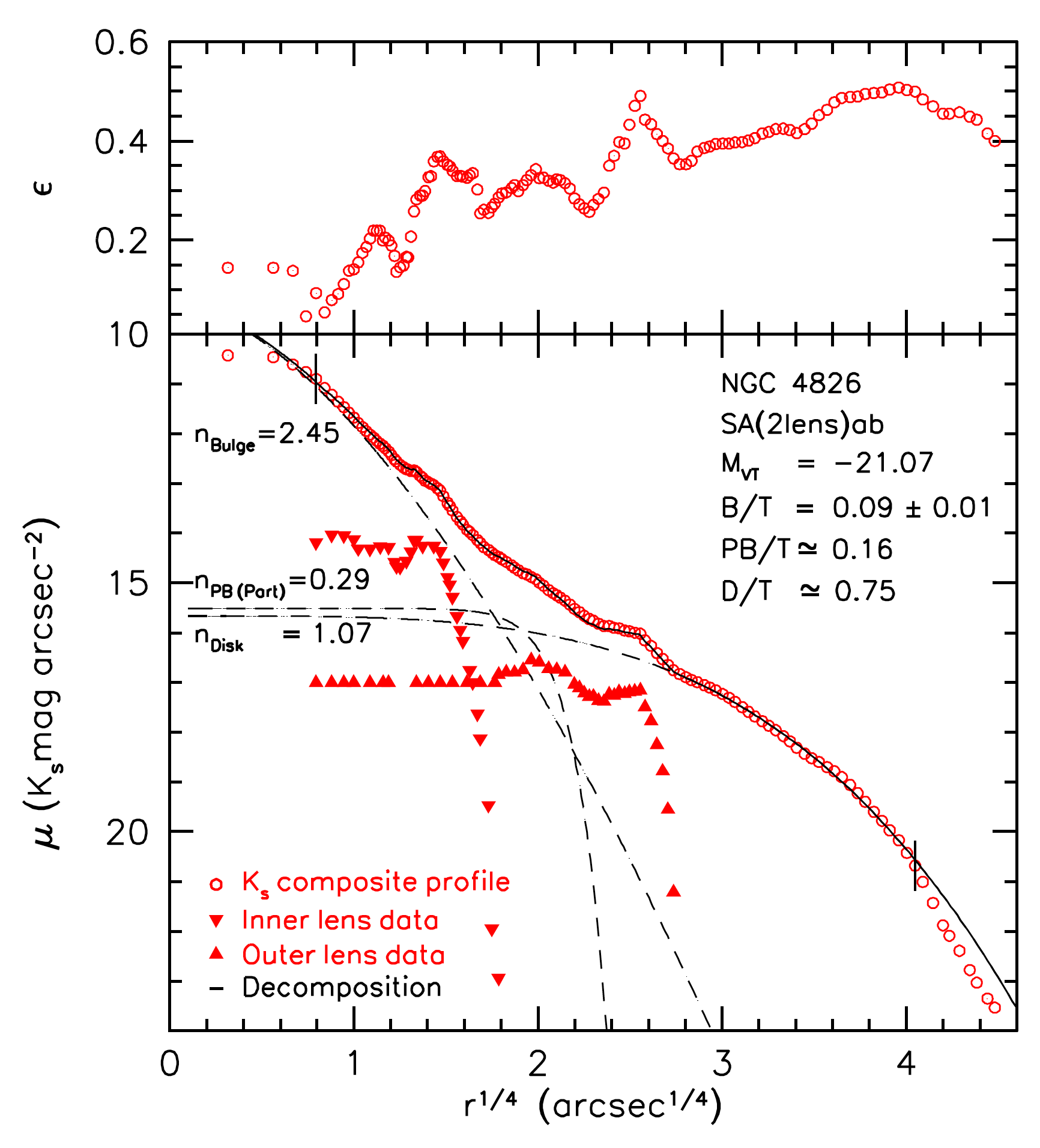}
\caption{
 Maximal-pseudobulge photometric decomposition plotted against $(r/\mathrm{arcsec})^{1/4}$ to show components 
 at both small and large radii.  In this fit, the  parameters were derived to maximize the pseudobulge-to-total
 ratio $PB/T$ while still producing an acceptable global fit.  This procedure required forcing the classical bulge S{\'e}rsic index to its lowest value.  The lens components are treated as they are in our adopted fit.    The classical bulge component contributes roughly half as much light as the pseudobulge, but it still dominates the light inside of $(r/\mathrm{arcsec})^{1/4} = 1.5$.  The low classical bulge to total ratio, while not ruled out, is small for for an Sab galaxy.  Data and symbols are the same as in Fig.\ \ref{fig:decomp1r4}.
\label{fig:decomp3r4}}
\end{figure}

\appendix
\label{appendix}
\hypertarget{appendixbeginning}Figures \ref{fig:decomp2r4} and \ref{fig:decomp3r4} illustrate our alternative photometric decompositions of the
brightness profile of NGC 4826. Fig.\ \ref{fig:decomp2r4} has the biggest allowed classical bulge:~the pseudobulge has been reduced to the two shelves in the brightness distribution, which reveal themselves 
to be disky via local maxima in elipticity $\epsilon$.  The rest of the inner component of a 2-S\'ersic fit 
is here called the classical bulge.  The fit to the profile is acceptable, but the decomposition is implausible,
 because the disk has a smaller S\'ersic index $n_{\rm Disk} = 0.87 \pm 0.04$ than do other disk galaxies 
 and the bulge has a larger S\'ersic index than do the bulges of spiral galaxies.  In fact, 
 $n_{\rm Bulge} = 4.28 \pm 0.24$ is characteristic of the S\'ersic indices $n > 4$ of giant elliptical 
 galaxies with cores (Kormendy et al.~2009).
 
In contrast, Figure \ref{fig:decomp3r4} is a fit in which the pseudobulge is constrained to be as luminous as possible.
This is done by forcing the classical bulge to have a small S\'ersic index, \hbox{although} still one that is 
consistent with other bulge-disk galaxies.  

Note that the maximal (Figure \ref{fig:decomp2r4}), adopted (Figures \ref{fig:decomp1r} \ref{fig:decomp1r4}), and minimal (Figure \ref{fig:decomp3r4}) classical bulges 
all dominate the light of the galaxy in the central few arcsec, i.e., inside and near the 
sphere-of-influence volume of the black hole.

\vfill

As discussed in Section \ref{obs:imag:decomp}, one test of our photometric component decomposition remains to be carried out.
Are the structural parameters of the classical bulge consistent with our understanding of other galaxies?
The ``fundamental plane'' parameter correlations between bulge luminosity, effective radius $r_e$, and
effective brightness $\mu_e$ are tight enough for classical bulges so that a meaningful test is possible.
In contrast, pseudobulges have larger parameter ranges and are less well studied and understood.  So no
meaningful test is possible of the pseudobulge parameters that we derive for NGC 4826.

\begin{figure}[htb]
\centering
\vspace{1 cm}
\includegraphics[trim=132 190 135 120 , width=\columnwidth]{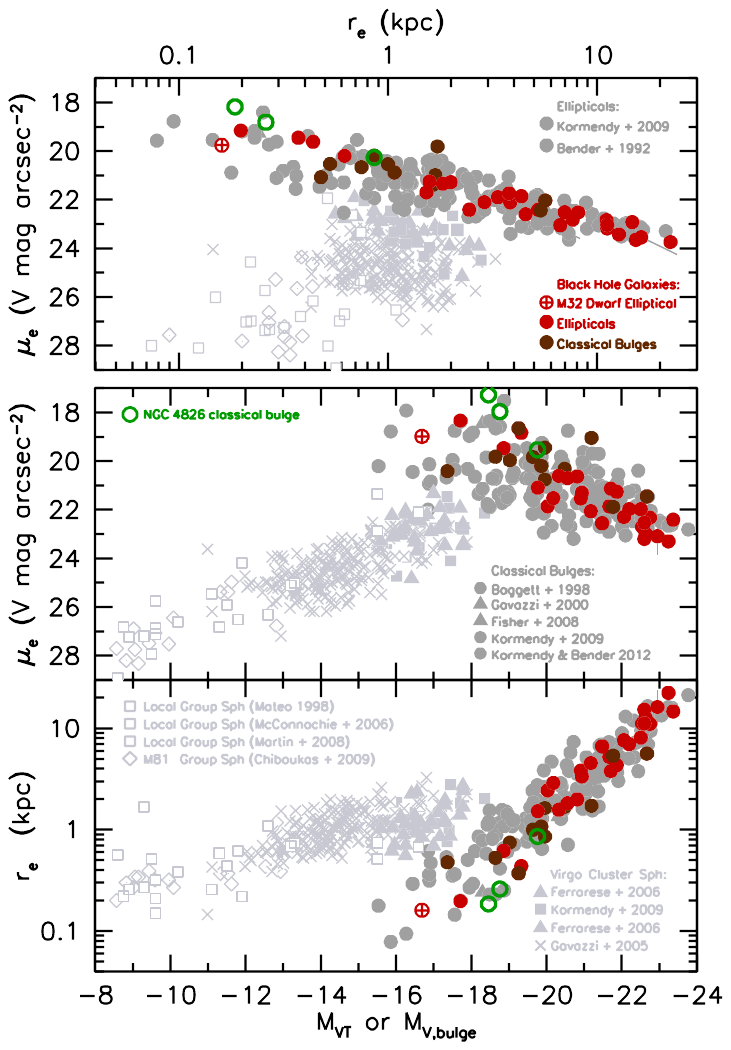}
\caption{
From \citet{2020IAUS..353..186K}, parameter correlations for elliptical galaxies and classical
bulges of disk galaxies (dark gray) and for spheroidal galaxies (light gray). The bottom panels show
effective radius $r_e$ that contains half of the total light and effective surface brightness $\mu_e$ at 
$r_e$ correlated with the $V$-band absolute magnitude of the component. The top panel shows the \citet{1977ApJ...218..333K} 
relation, $\mu_e$ versus $r_e$; this projection shows the fundamental plane nearly edge-on and has especially
small scatter. Sources are given in keys, references omitted here are in \citet{2012ApJS..198....2K}.
Galaxies in which supermassive black holes are detected via spatially resolved stellar or gas dynamics
are encoded in dark red for ellipticals and dark brown for bulges. Our three classical bulge models for
NGC 4826 are shown in green; the middle point is for the preferred decomposition.
\label{fig:FP}}
\end{figure}

\begin{figure}[htb]
\centering
\includegraphics[trim=140 245 60 210 , width=0.9\columnwidth]{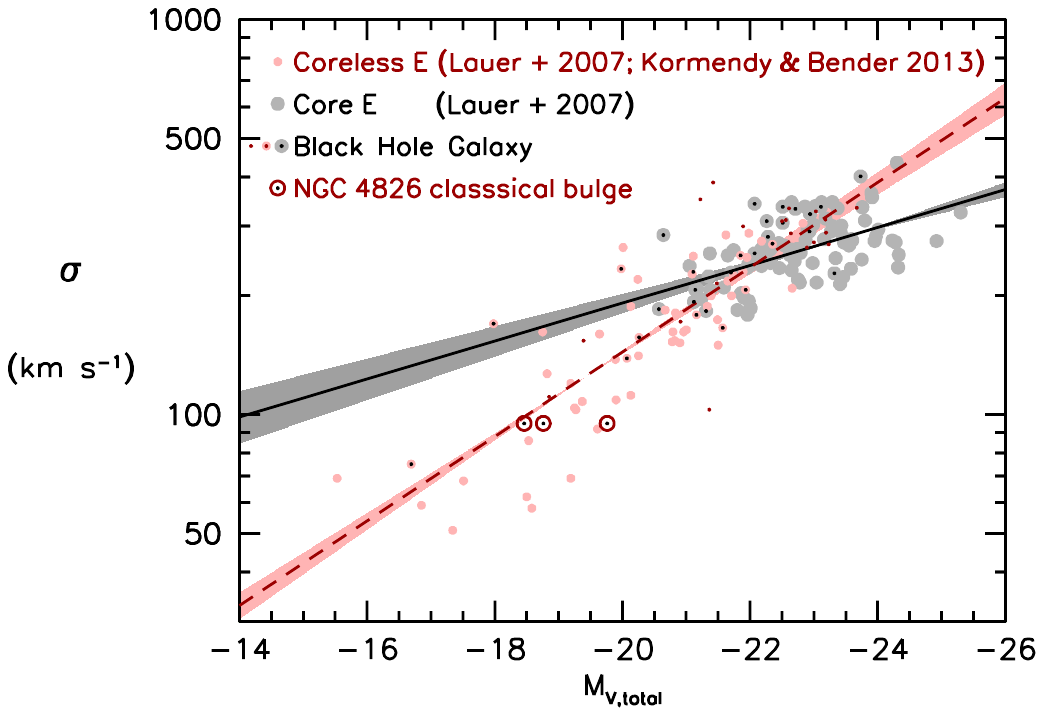}
\caption{
\citet{1976ApJ...204..668F}  correlations for ellipticals with and without a ``core'', i.e., a break
in the density profile near the center from a steep outer profile to a shallow inner cusp. Total absolute
magnitudes and velocity dispersions are from \citet{2007ApJ...662..808L}  with classification corrections from
\citet{2009ApJS..182..216K} and \citet{2013ApJ...769L...5K}. Galaxies with dynamical black hole detections are
labeled. The lines are symmetric least-squares fits \citep{2002ApJ...574..740T} to the core galaxies (black
line with gray $1\sigma$ uncertainty) and the coreless galaxies (red line with pink $1\sigma$ uncertainty). 
The kink at $\sigma \simeq 250$ km s$^{-1}$ was also emphasized by \citet{2007ApJ...662..808L}. This figure is 
from \citet{2020IAUS..353..186K} with the three classical bulge models for NGC 4826 added. All are consistent with the 
correlation for coreless ellipticals.  The smaller bulges are slightly favored.
\label{fig:FJ}}
\end{figure}

Two structural parameter correlations from \citet{2020IAUS..353..186K} are shown in Figures \ref{fig:FP} and \ref{fig:FJ} with the
classical bulge parameters of NGC 4826 added.  Figure \ref{fig:FJ} uses data and classifications due to \citet{2007ApJ...662..808L}, 
\citet{2009ApJS..182..216K}, \citet{2013ApJ...769L...5K}.  \citet{2020IAUS..353..186K} showed that the structural parameters of 
those classical bulges and elliptical galaxies in which black holes have been detected via spatially 
resolved dynamics are entirely normal and sample the complete range of properties observed for classical
bulges and ellipticals. This argues against claims \citep[e.g.,][]{2015ApJS..218...10V, 2016ApJ...831..134V, 2016MNRAS.460.3119S} 
that observed correlations between black hole mass and bulge properties are biased because black holes
are detected only in a biased subset of galaxies that have especially favorable properties.  The only notable exceptions to this are galaxies with $r_e \lesssim 1\ \units{kpc}$ have black hole mass measurements only for brighter values of $\mu_e$, galaxies with $M_{V,\mathrm{bulge}}$ or $M_{VT} > -20$ have black hole mass measurements for brighter values of $\mu_e$ and/or smaller values of $r_e$.

Here, we see that the bulge models of NGC 4826 lie within the scatter of the correlations for
bulges and elliptical galaxies. Actually, the bulge of NGC 4826 is more compact than typical bulges.
The smallest bulge is weakly disfavored by the Figure \ref{fig:FP} correlations. But overall, the bulge of this galaxy
is essentially normal for its $M_V$, in terms of the correlations between $r_e$, $\mu_e$, and $M_V$.

Note that a black hole is successfully detected in NGC 4826, at the compact extreme of the distribution
of bulge properties, and in M32, more nearly at the diffuse side of the scatter of bulge properties, 
both at the low-luminosity end of the parameter correlations.

The same is true for the Faber-Jackson (1976) correlation between $M_V$ and the velocity dispersion
$\sigma_{*}$ outside the sphere-of-influence radius of the black hole. Core and coreless galaxies have different
Faber-Jackson relations (Lauer et al.~2007; Kormendy \& Bender 2013), just as they have different
S\'ersic indices (e. g., Kormendy et al.~2009). But all three bulge models for NGC 4826 are consistent
with the Faber-Jackson relation for similar-luminosity coreless ellipticals. The largest-bulge  
model is slightly disfavored but not excluded.

% ORCID iDs

% Kayhan G\"ultekin    https://orcid.org/0000-0002-1146-0198

% Karl Gebhardt        https://orcid.org/0000-0002-8433-8185

% John Kormendy        https://orcid.org/0000-0001-9854-5217

% Adi Foord            https://orcid.org/0000-0002-1616-1701

% Ralf Bender          https://orcid.org/0000-0001-7179-0626

% Tod Lauer            https://orcid.org/0000-0003-3234-7247

% Scott Tremaine       https://orcid.org/0000-0002-0278-7180

\bibliography{gultekin}{}
\bibliographystyle{aasjournal}

\end{document}